\documentclass{aa}
\usepackage{graphicx}
\def\H2{H$_2$}
\def\fH2{f_{\rm H_2}}
\def\tff{t_{\rm ff}}
\def\izw{I\,Zw\,18}
\def\sbs{SBS\,0335$-$052}
\begin{document}
\title{The role of dust in ``active'' and ``passive'' low-metallicity
star formation}
\author{H. Hirashita \inst{1}
\fnmsep\thanks{Research Fellow of the Japan Society for the Promotion of
Science.}
         \and
L. K. Hunt \inst{2}
}
\offprints{H. Hirashita,\\ \email{hirashita@u.phys.nagoya-u.ac.jp}}
\institute{Graduate School of Science, Nagoya University, Furo-cho,
                Chikusa-ku, Nagoya, 464-8602, Japan\\
\email{hirashita@u.phys.nagoya-u.ac.jp}
        \and
Istituto di Radioastronomia-Sezione Firenze, Largo E. Fermi, 5,
50125 Firenze, Italy\\
\email{hunt@arcetri.astro.it}
}
\date{Recieved 10 December 2003 / Accepted 1 April 2004}
\abstract{
We investigate the role of dust in star formation activity of
extremely metal-poor blue compact dwarf galaxies (BCDs).
Observations suggest that star formation in BCDs occurs in two
different regimes: ``active'' and ``passive''. The ``active''
BCDs host super star clusters (SSCs), and are characterised by
compact size, rich \H2 content, large dust optical depth, and
high dust temperature; the ``passive'' BCDs are more diffuse with
cooler dust, and lack SSCs and large amounts of \H2. By treating
physical processes concerning formation of stars and dust, we are
able to simultaneously reproduce all the above properties of both
modes of star formation (active and passive). We find that the
difference between the two regimes can be understood through the
variation of the ``compactness'' of the star-forming region:
an ``active'' mode emerges if the region is compact
(with radius $\la 50$ pc) and dense (with gas number density
$\ga 500$ cm$^{-3}$). The dust, supplied from Type II
supernovae in a compact star-forming region, effectively
reprocesses the heating photons into the infrared and induces a
rapid \H2 formation over a period of several Myr. This explains the
high infrared luminosity, high dust temperature, and large \H2
content of active BCDs. Moreover, the gas in ``active''
galaxies cools ($\la 300$ K) on a few dynamical timescales,
producing a ``run-away'' star formation episode  because of the
favourable (cool) conditions.
The mild extinction and relatively low molecular
content of passive BCDs can also be explained by the same
model if we assume a diffuse region (with radius
$\ga 100$ pc and gas number density $\la 100$ cm$^{-3}$). We
finally discuss primordial star formation in high-redshift
galaxies in the context of the ``active'' and ``passive'' star
formation scenario.
 }
\titlerunning{Dust in blue compact dwarf galaxies}
\authorrunning{H. Hirashita and L. K. Hunt}
\maketitle
%


\section{Introduction}\label{sec:intro}

The surfaces of interstellar dust grains are known to be sites
where efficient formation of hydrogen molecules (\H2) takes place.
Without dust grains, \H2 forms only in the gas phase with a
production rate much lower than the dust surface reaction
(e.g., Peebles \& Dicke \cite{peebles68};
Matsuda et al.\ \cite{matsuda69}). The shielding of \H2
dissociating photons by dust grains also enhances molecule formation.
In general, dust also absorbs a part of the radiation from stars
and reemits it in the infrared (IR)\footnote{In this paper, IR
indicates the wavelength range where the emission from dust
dominates the radiative energy (roughly 8--1000 $\mu$m).},
thereby modifying the spectral energy distribution (SED) of
galaxies (e.g., Silva et al.\ \cite{silva98}). Therefore, dust
plays an important role in both chemical and radiative properties
of galaxies.

Recently Hirashita \& Ferrara (\cite{hirashita02}, hereafter
HF02) have proposed that the dust enrichment in extremely
metal-poor primeval objects is essential for the enhancement
of star formation activity. Their scenario suggests the
following cycle between dust production and star formation:
{\it (i)}~massive stars end their lives as Type II supernovae
(SNe II), which supply dust grains into the interstellar
medium (ISM) (Kozasa et al.\ \cite{kozasa89};
Todini \& Ferrara \cite{todini01};
Nozawa et al.\ \cite{nozawa03};
Schneider et al.\ \cite{raffa04}); {\it (ii)}~those grains enhance
the formation of molecular clouds in which stars form;
{\it (iii)}~some of those stars are massive and the dust supply
described in {\it (i)} occurs again. This cycle {\it (i)}--{\it (iii)} 
effectively enhances the star formation rate as shown by HF02.

In fact, a large amount of dust has been suggested to exist at high
redshift (high $z$) (e.g., Smail et al.\ \cite{smail97}).
However, it is not easy to explore the first dust enrichment in
primeval galaxies at high $z$ ($z\ga 5$) with the present
observational facilities. Therefore, nearby templates for primeval
galaxies are useful to test galaxy formation scenarios. The best
candidates for such a template are metal-poor blue compact dwarf
galaxies (BCDs),
since they are at the initial stage of metal enrichment and
their current star formation activity is
generally young (Searle \& Sargent \cite{searle72};
Kunth \& \"{O}stlin \cite{kunth00}). In other words, BCDs
can be used as laboratories in which to study high-$z$ primeval galaxies.

Two classes of star formation activity have recently emerged
observationally, as proposed for a BCD sample by
Hunt et al.\ (\cite{hunt-cozumel}, hereafter HHTIV;
\cite{hunt-active}). They argue that the star-formation activity in
the two most metal-poor galaxies, \sbs\ and \izw, shows very different
properties, in spite of their similar metallicities
(1/41 and 1/50 $Z_\odot$, respectively;
Skillman \& Kennicutt \cite{skillman93};
Izotov et al.\ \cite{izotovetal99}). The major star-forming
region of \sbs\ is compact and dense 
(radius $r_{\rm SF}\la 40$ pc, number density $n\ga 600$ cm$^{-3}$;
Dale et al.\ \cite{dale01}; Izotov \& Thuan \cite{izotov99}).
Moreover, \sbs\ hosts
several super star clusters (SSCs), detectable
\H2 emission lines in the near-infrared
(Vanzi et al.\ \cite{vanzi00}), a large dust extinction
($A_V\sim 16$ mag; Thuan et al.\ \cite{thuan99};
Hunt et al.\ \cite{hunt01};
Plante \& Sauvage \cite{plante02}), and high dust temperature
(Hunt et al.\ \cite{hunt01}; Dale et al.\ \cite{dale01};
Takeuchi et al.\ \cite{takeuchi03}). On the contrary, the
star-forming regions in \izw\ are diffuse
($r_{\rm SF}\ga 100$ pc, $n\la 100$ cm$^{-3}$), and contain
no SSCs. Near-infrared \H2 emission has not been detected (Hunt et al.,
private communication), and the dust extinction 
is moderate ($A_V\sim 0.2$ mag; Cannon et al.\ \cite{cannon02}). We
call a region with such properties ``passive'' following HHTIV.
The similar metallicities of \sbs\ (active) and
\izw\ (passive) imply that
the chemical abundance is not a primary factor in determining the
star-forming properties. We argue that the compactness of
star-forming regions, which affect gas density,
gas dynamics, and so on, is important in the dichotomy
of active and passive modes.

Hirashita et al.\ \cite{hhf02} (hereafter HHF02) show that the IR
luminosity, the dust mass, and the rich \H2 content of \sbs\
can be explained through dust accumulation by successive SNe II
in a compact ($r_{\rm SF}\la 100$ pc) region. Moreover, \sbs\ is
not unique among BCDs; the BCDs with dense
compact star-forming regions similar to \sbs\ were dubbed ``active'',
and tend to be characterized by high surface brightness
(see Figure 1 of HHTIV). The physical state of ISM is also
similar among ``active'' BCDs. They always have large dust
extinction, and a significant fraction of stellar radiation is
reprocessed into IR; the dust temperature is also high and
there is a ``hot'' component with 600--1000 K
(Hunt et al.\ \cite{hunt02}). Dust properties can be further
constrained by future IR observations for a large sample of
BCDs (e.g., Takeuchi et al.\ \cite{takeuchi03}). HHTIV also
demonstrate that there are BCDs with converse properties, namely
``passive'' ones, which are more diffuse, less dense,
and of lower surface brightness.
A representative of this category is \izw. 
Contrary to ``active'' BCDs, IR dust
emission has not been detected so far in ``passive'' BCDs.
Although their star formation rate is
lower than ``active'' BCDs, ``passive'' BCDs 
are actually forming stars, 
and are completely different from passively evolving
galaxies such as ellipticals.

The above clear difference in dust properties implies that
in addition to the compactness, dust should be considered as 
a key to understand the ``active'' and
``passive'' modes, and hence to understanding how star formation
occurs in extremely low-metallicity environments.
We consider the role of dust in various compactness of regions by using our
theoretical framework.
This paper is organised as follows. First, in
Section \ref{sec:model} we explain the model that describes
the evolution of dust content and gas state in a star-forming
region. Then, in Section \ref{sec:result}, we present our
results concerning the differences between ``active'' and
``passive'' star-forming regions. In
Section \ref{sec:bcds}, we discuss the interpretation
of observational properties of active and passive BCDs in the
context of our model, and consider in particular the two 
prototypes, \izw\ and \sbs. 
Implications for high-redshift star formation are described in
Section \ref{sec:highz}, and our conclusions are presented in
Section \ref{sec:conclusion}.

\section{Model description}\label{sec:model}

For our calculation of dust amount in a star-forming region,
we essentially use the model by HF02. We treat the chemical
reaction network concerning \H2 formation in a way consistent
with dust amount evolution. We approximate star-forming
regions as homogeneous spheres and adopt a representative
value for each physical quantity (i.e., we treat a
star-forming region as one zone).

\subsection{ISM evolution}\label{subsec:phys_state}

The star formation process is affected by the physical state of
the ISM. In metal-poor environments, cooling by molecular hydrogen
plays an important role in star formation
(Galli \& Palla \cite{galli98}; Abel et al.\ \cite{abel00};
Bromm et al.\ \cite{bromm01}; Nishi et al.\ \cite{nishi98};
Omukai \cite{omukai00}; Kamaya \& Silk \cite{kamaya02};
Ripamonti et al.\ \cite{ripamonti02}). The abundance of \H2 should be
ultimately a key parameter since stars are only seen to form in
molecular complexes (see e.g., Wilson et al.\ \cite{wilson00}
for a recent observation of a nearby galaxy). Moreover,
enhancement of molecular gas formation is shown to result in
an active star formation (for recent results, see e.g.,
Walter et al.\ \cite{walter02}), and indeed there is a
correlation between molecular amount and star formation rate
(e.g., Bendo et al.\ \cite{bendo02}). We define the molecular
fraction, $\fH2$, as
\begin{eqnarray}
\fH2\equiv 2n_{\rm H_2}/n_{\rm H}\, ,\label{eq:mol_frac}
\end{eqnarray}
where $n_{\rm H}$ and  $n_{\rm H_2}$ are the number densities
of hydrogen nuclei and hydrogen molecules, respectively (i.e.,
if all the hydrogen nuclei are in the molecular form, $\fH2 =1$).

We then calculate the time evolution of ionisation degree ($x$),
molecular fraction ($\fH2$), and gas temperature ($T$) of the
hydrogen gas. The helium content is neglected here, since its effect 
should be negligible for the three quantities
(Kitayama et al.\ \cite{kitayama01}; HF02). Therefore, the
number density of the gas ($n$) is approximated by the hydrogen
number density (i.e., $n\simeq n_{\rm H}$). The ionisation degree
and the temperature affect the formation rate of \H2.
The processes are treated as follows (see HF02 for details).

We calculate the time evolution of the ionisation degree $x$ by taking 
into account collisional ionisation, recombination, and
photo-ionisation of hydrogen. For the photo-ionisation, we have
included the radiative
transfer effect following the simple recipe in Appendix of
Kitayama \& Ikeuchi (\cite{kitayama01}), who derive an analytical
expression under the assumption of a power law spectrum of
incident photons. 
For the spectral shape parameter
$\alpha$ of the radiation field, we adopt $\alpha =3$, because
the resulting ratio between \H2 destroying photons and
H ionizing photons is similar to realistic OB stars.
Kitayama et al.\ (\cite{kitayama01}) have examined
$\alpha =1$ and $\alpha =5$. Their table 1 indicates that
$\alpha =1$ overproduces the ionizing photons while
$\alpha=5$ overproduces the \H2 destroying photons. Therefore,
we adopt an intermediate value: $\alpha =3$. 
In any case, different values of $\alpha$ do not affect the behaviour 
of $x$, $\fH2$, and $T$ as functions of time, because the dust
accumulation (which does not depend on $\alpha$) is the greatest influence
on these three quantities. The normalisation of the intensity,
$I_0(\nu_{\rm HI})$, is determined from
\begin{eqnarray}
cu_{\rm UV}\equiv\frac{L_{\rm UV}}{(r_{\rm SF}/2)^2}=4\pi
\int_{\nu_{\rm min}}^\infty I_0(\nu_{\rm HI})\left(
\frac{\nu}{\nu_{\rm HI}}\right)^{-\alpha}\,d\nu \ , \label{eq:isrf}
\end{eqnarray}
where $\nu_{\rm min}(\simeq 10^{15}~{\rm Hz})$ is the minimum
frequency where OB stars dominate the radiative energy, $u_{\rm UV}$
is the ``typical'' interstellar radiation field estimated at the
half of the radius ($r_{\rm SF}/2$; half is for a rough average),
$\nu_{\rm HI}$ is the ionisation frequency of neutral hydrogen
($\nu_{\rm HI}=3.3\times 10^{15}$ Hz) (see HF02 for details), and
$c$ is the light speed.

In calculating $\fH2$, we consider \H2 formation both
in the gas phase (via H$^-$ or H$_2^+$, where the latter is
negligible), and on the dust grain surface; \H2 destruction occurs
through collisions with H$^+$, H, and e$^-$, and photodissociation.
For self-shielding effects which prohibit \H2 dissociation,
equation (17) of HF02 is used, but here we substitute
$r_{\rm disc}$ with $r_{\rm SF}/2$. For the reaction rates, we
adopt table 1 of HF02, but for the reaction on dust grains,
$R_{\rm dust}$, we use the following expression which includes
the dependence on $a$:
\begin{eqnarray}
R_{\rm dust}=\left\{
  \begin{array}{l}
    2.8\times 10^{-15}\left(
    \frac{\displaystyle T}{\displaystyle 100~{\rm K}}\right)^{1/2}\left(
    \frac{\displaystyle a}{\displaystyle 0.03~\mu{\rm m}}\right)^{-1} \\
    {~~}\times\left(
    \frac{\displaystyle \delta}{\displaystyle 2~{\rm g}~{\rm cm}^{-3}}
    \right)^{-1}~{\rm cm}^3~{\rm s}^{-1} ~ \mbox{if $T\leq 300$ K}\\
	\\
    0 ~~~ \mbox{if $T>300$ K}
  \end{array}
\right.\label{eq:reaction_h2}
\end{eqnarray}

In order to calculate the temperature evolution, cooling and heating
must be considered in our model. We assume that cooling processes
comprise collisional excitation and ionisation of atomic hydrogen
(when $T\ga 8000$ K) and collisional excitation of molecular
hydrogen (which dominates for $T< 8000$ K). For the heating by stellar
UV radiation, we have included the radiative transfer effect by
following a simple analytical recipe in Appendix of
Kitayama \& Ikeuchi (\cite{kitayama00}).

\subsection{Star formation rate}\label{subsec:sfr}

Stars form as a result of the gravitational collapse of a gas
cloud. Therefore, it is physically reasonable to relate the star
formation rate with the free-fall timescale of gas. We consider
a star-forming region with a gas number density of
$n\sim n_{\rm H}$. The free-fall time, $\tff $, is estimated as
\begin{eqnarray}
\tff \simeq\frac{1}{\sqrt{G\rho\,}\,}\simeq 1\times 10^7
\left(\frac{n_{\rm H}}{100~{\rm cm}^{-3}}
\right)^{-1/2}~{\rm yr}\, ,
\label{eq:freefall}
\end{eqnarray}
where $\rho = m_{\rm H}n_{\rm H}$ ($m_{\rm H}$ is the mass of a
hydrogen atom) is the mass density of the gas. The star
formation rate, $\psi$, is estimated as
\begin{eqnarray}
\psi & = & \frac{\epsilon_{\rm SF}M_{\rm gas}}{\tff }\,
\Theta (t)\nonumber\\
& \simeq & 0.1\left(\frac{\epsilon_{\rm SF}}{0.1}\right)
\left(\frac{M_{\rm gas}}{10^7~M_\odot}\right)\left(
\frac{n_{\rm H}}{100~{\rm cm}^{-3}}
\right)^{1/2}\Theta (t)\nonumber\\
& & M_\odot~{\rm yr}^{-1}\, ,
\label{eq:sfr}
\end{eqnarray}
where $M_{\rm gas}$ is the total gas mass (both molecular
and atomic) of the star-forming region, $\epsilon_{\rm SF}$
is the star formation efficiency defined as the conversion
efficiency of a gas into stars over a free-fall time, and
$\Theta (t)$ is Heaviside's step function [$\Theta (t)=1$
if $t\geq 0$ and
$\Theta (t)=0$ if $t<0$]. Thus, we define the zero point of
time $t$ at the onset of star formation in the star-forming
region. For simplicity, we assume a constant star formation
rate (i.e., $\epsilon_{\rm SF}$, $M_{\rm gas}$, and $n_{\rm H}$
are approximated to be constant) and a spherical homogeneous
star forming region. In reality, there is probably a
significant density inhomogeneity; however,
the star formation rate in the entire star-forming region may
be regulated by the dynamical time governed by a
spatially averaged density, and the
homogeneous density can be regarded as such an average.

The hydrogen number density can be related
to gas mass as
\begin{eqnarray}
\frac{4\pi}{3}r_{\rm SF}^3n_{\rm H}m_{\rm H}=M_{\rm gas}\, ,
\end{eqnarray}
where $r_{\rm SF}$ is the radius of the star-forming region.
Thus, the numerical value of the number density is estimated as
\begin{eqnarray}
n_{\rm H}\simeq 100 \left(\frac{r_{\rm SF}}{100~{\rm pc}}
\right)^{-3}\left(\frac{M_{\rm gas}}{10^7~M_\odot}
\right)~{\rm cm}^{-3}\, .\label{eq:density}
\end{eqnarray}
We also define the gas consumption timescale $t_{\rm gas}$:
\begin{eqnarray}
t_{\rm gas}\equiv\frac{M_{\rm gas}}{\psi}=
\frac{\tff}{\epsilon_{\rm SF}}\, .
\label{eq:tgas}
\end{eqnarray}
Since the effect of gas conversion into stars becomes
significant for $t\sim t_{\rm gas}$, we stop the calculation
at $t=t_{\rm gas}/2$. After this time, star formation
may be suppressed because of the lack of gas.

\subsection{Evolution of dust content}
\label{subsec:dust_formalism}

We calculate the evolution of dust mass in a star-forming region
considering the dust supply from SNe II. The preexisting dust at
$t=0$ in the star-forming region is neglected here.
This is a good approximation for the extremely metal-poor BCDs 
treated in this paper. In fact, a large fraction of dust amount contained
in some BCDs can be explained by the present star formation activity
(see Section \ref{subsec:chemical}). Preexisting dust would strengthen all
the dust effects described below.

If the starburst is young, and there has been no previous burst, 
dust can be supplied only from massive stars with short lifetimes.
Thus, SNe II are the dominant source of dust formation if the
age is typically less than a few times $10^8$ yr
(Dwek \cite{dwek98}).
 Thus, the
dust formation rate is related to the rate of SNe II. The SN II
rate as a function of time, $\gamma (t)$, is given by
\begin{eqnarray}
\gamma (t)=\int_{8~M_\odot}^{\infty}\psi (t-\tau_m)\, \phi (m)\,
dm\, ,\label{eq:sn2}
\end{eqnarray}
where $\psi (t)$ is the star formation rate at $t$
(Eq.\ \ref{eq:sfr}), $\phi (m)$ is the initial mass function
(IMF; the definition of the IMF is the same as that in
Tinsley \cite{tinsley80}), $\tau_m$ is the lifetime of a star
whose mass is $m$, and we assume that only stars with $m>8~M_\odot$
produce SNe II. In this paper, we assume a Salpeter IMF
($\phi (m)\propto m^{-2.35}$) with the stellar mass range of
0.1--60 $M_\odot$ (HF02). If we assume a higher upper limit
such as 100 $M_\odot$, we expect a higher SN II rate and dust
production rate. However, if the progenitor mass $m$ is larger
than $\sim 50~M_\odot$, almost all the produced metals may collapse
into a central blackhole (Tsujimoto et al.\ \cite{tsujimoto97}).
Todini \& Ferrara (\cite{todini01}) treated only 
progenitor stellar mass less than 40 $M_\odot$, and the
dust production rate for massive stars ($\ga 60~M_\odot$)
is unknown theoretically. Recently, however, there have been some
advances in this field (Nozawa et al.\ \cite{nozawa03};
Schneider et al.\ \cite{raffa04}), and we are
developing a model to include the effect of massive stars.

Dust destruction by shocks from SNe II can be important
(Dwek \& Scalo \cite{dwek80}; Jones et al.\ \cite{jones96}). The
destruction timescale $\tau_{\rm SN}$ is estimated to be
(McKee \cite{mckee89}; Lisenfeld \& Ferrara \cite{lisenfeld98})
\begin{eqnarray}
\tau_{\rm SN}=
\frac{M_{\rm gas}}{\gamma\epsilon_{\rm SN}M_{\rm s}}\, ,
\end{eqnarray}
where $M_{\rm s}$ is the mass accelerated to a velocity large
enough for dust destruction by a SN blast ($\sim 100$~km~s$^{-1}$),
$\gamma$ is the SN II rate, $\epsilon_{\rm SN}$ is the efficiency
of dust destruction in a medium shocked by a SN II. We adopt
$M_{\rm s}=6.8\times 10^3~M_\odot$
(Lisenfeld \& Ferrara \cite{lisenfeld98}; see also
Tielens \cite{tielens98}) and $\epsilon_{\rm SN}=0.1$
(McKee \cite{mckee89}). Since we are interested in extremely
metal-poor environments, we assume the relation between stellar
mass and lifetime of zero-metallicity stars in table 6 of
Schaerer (\cite{schaerer02}).\footnote{The case without mass loss is
applied for consistency with HF02. This does not unduly affect our
results however since the differences in metallicity and
luminosity for the case {\it with} and {\it without} mass loss are
within a factor of 2.} 

Dust grains can also be destroyed 
by strong and hard interstellar radiation fields
(e.g., Boulanger et al. \cite{boulanger88}; Puget \& L{\' e}ger \cite{puget89};
Voit \cite{voit92}; Contursi et al. \cite{contursi00}).
However, the grain populations most strongly affected are
the carriers of the aromatic band features 
(e.g., Polycyclic Aromatic Hydrocarbons) and very small grains 
(diameter $a<0.01\mu$m), neither of which are considered here.
Hence, since grain destruction processes by UV radiation
are most severe for very small grains, 
and in any case are poorly known, we will neglect them.

In this case the rate of increase of $M_{\rm dust}$ is written as
\begin{eqnarray}
\frac{dM_{\rm dust}}{dt}=m_{\rm dust}\gamma-
\frac{{M}_{\rm dust}}{\tau_{\rm SN}}\, ,\label{eq:dust}
\end{eqnarray}
where $m_{\rm dust}$ is the typical dust mass produced in a
SN II. Todini \& Ferrara (\cite{todini01}) show that
$m_{\rm dust}$ varies with progenitor mass, metallicity ($Z$),
and input energy of SNe II. The Salpeter IMF-weighted
mean of dust mass produced per SN II for (1) $Z=0$, Case A,
(2) $Z=0$, Case B, (3) $Z=10^{-2}~Z_\odot$, Case A, and
(4) $Z=10^{-2}~Z_\odot$, Case B are (1) 0.22 $M_\odot$,
(2) 0.46 $M_\odot$, (3) 0.45 $M_\odot$, and (4) 0.63 $M_\odot$,
respectively ($Z$ is the metallicity, and Cases A and B
correspond to low and high explosion
energy\footnote{The kinetic energies given to the ejecta
are $1.2\times 10^{51}$ erg and $2\times 10^{51}$ erg} for
Case A and Case B, respectively.)
We adopt the average of the four cases, i.e.,
$m_{\rm dust}=0.4~M_\odot$ (HHF02). 
This may not be an overly optimistic estimate given the low upper 
mass cutoff of 35\,$M_\odot$ adopted by Todini \& Ferrara
(\cite{todini01}), and the higher dust production estimates
of Nozawa et al. (\cite{nozawa03}).
The calculated dust mass
$M_{\rm dust}$ is roughly proportional to $m_{\rm dust}$,
and the timescale on which the dust extinction effects appear
is approximately proportional to $1/m_{\rm dust}$.

\subsection{Evolution of metal content}\label{subsec:metal}

The evolution of metal content can be calculated once the
star formation history and metal yield are fixed (e.g.,
Tinsley \cite{tinsley80}). Because of the young age range
($\la 10^8$ yr) treated in this paper, we assume that the metal
production is dominated by SNe II. The evolution of the mass of
metals (a given species is indicated by $i$) in the gas phase,
$M_i$, is calculated by subtracting the dust mass from the metal
mass. This is expressed as
\begin{eqnarray}
\frac{dM_i}{dt}=m_i\gamma -\frac{dM_{{\rm dust},\,i}}{dt}\, ,
\label{eq:metal}
\end{eqnarray}
where $m_i$ is the IMF-weighted average of metal mass (for
species $i$) ejected per SN II and $M_{{\rm dust},\,i}$ is
the mass of element $i$ in dust phase. We calculate the
abundances of carbon and oxygen, because those two elements
are the two major metals produced in SNe II
(Woosley \& Weaver \cite{woosley95}). Following HF02, we
adopt $M_{\rm dust,\,O}=0.15M_{\rm dust}$,
$M_{\rm dust,\,C}=0.36M_{\rm dust}$, $m_{\rm O}=1.2~M_\odot$,
and $m_{\rm C}=0.17~M_\odot$; i.e., we assume that
carbonaceous grains are responsible for 36\% of the total dust
mass, and the other grains contain oxygen with mass fraction of
23\% (i.e., the oxygen fraction in mass is $0.64\times 0.23
=0.15$).

\subsection{Radiative properties}\label{subsec:luminosities}

One of the most direct ways to constrain dust content is to
observe IR continuum emission from dust grains. We now derive
the evolution of IR luminosity. Because of the large cross
section of grains against UV light and the intense UV
interstellar radiation field in star-forming galaxies, we
can assume that the IR luminosity is equal to the absorbed
energy of UV light.

\subsubsection{UV and IR}\label{subsubsec:UV_IR}

First, we estimate the fraction of the UV radiation absorbed
by dust. We define the following typical optical depth,
$\tau_{\rm dust}$, as
\begin{eqnarray}
\tau_{\rm dust}\equiv\pi a^2Q_{\rm UV}n_{\rm dust}r_{\rm SF}/2\, .
\label{eq:def_tau0}
\end{eqnarray}
where $\pi a^2Q_{\rm UV}$ is the typical absorption cross-section
for UV light ($Q_{\rm UV}$ is the dimensionless absorption
cross-section normalised by the geometrical cross-section), $a$
is the dust radius, $n_{\rm dust}$ is the mean dust grain
number density, and we assume that the typical path length
for a UV photon is half of $r_{\rm SF}$ as a rough
average.\footnote{In order to fully derive the precise path
length instead of roughly dividing $r_{\rm SF}$ by 2, a
detailed modelling of radiative transfer including dust
scattering is necessary. Here we only mention
that with a given dust mass, $\tau_{\rm dust}$ linearly
scales with the adopted path length.}
In this paper we assume for simplicity a single value for $a$,
consistently with the typical size of dust grains produced by
SNe II (e.g., Todini \& Ferrara \cite{todini01}; 
Nozawa et al.\ \cite{nozawa03}). The dust mass density $\delta$
is related to the mean dust number density $n_{\rm dust}$ as
\begin{eqnarray}
\frac{4\pi}{3}a^3\delta n_{\rm dust}\frac{4\pi}{3}r_{\rm SF}^3=
M_{\rm dust}\, ,\label{eq:dust_density}
\end{eqnarray}
where $\delta$ is the grain material density. By solving
Eq.~(\ref{eq:dust_density}) for $n_{\rm dust}$ and
substituting it into Eq.~(\ref{eq:def_tau0}), we obtain
\begin{eqnarray}
\tau_{\rm dust}=\frac{9}{32\pi}
\frac{Q_{\rm UV}M_{\rm dust}}{a\delta r_{\rm SF}^2}\, .
\end{eqnarray}

We define the ``attenuation function'' $E(\tau_{\rm dust})$ as
the fraction of UV light escaping from the star-forming
region. Thus, we write the UV luminosity of the star-forming
region as
\begin{eqnarray}
L_{\rm UV}\simeq L_{\rm UV,0}E(\tau_{\rm dust})\, ,
\label{eq:L_UV}
\end{eqnarray}
where $L_{\rm UV,0}$ is the intrinsic UV luminosity.
The functional form of $E(\tau_{\rm dust})$ depends on the
geometry of dust distribution, and we examine the following
two representative simple cases. One is the ``screen''
distribution, in which all the dust is located at the radius of
$r_{\rm SF}$ from the centre of the star-forming region.
The other is the ``mixed'' geometry, in which the spatial
distributions of dust and stars are the same (i.e.,
the mass ratio between stars and dust is constant). In the
screen geometry,
\begin{eqnarray}
E(\tau_{\rm dust})=E_{\rm sc}(\tau_{\rm dust})\equiv
\exp (-\tau_{\rm dust})\, ,\label{eq:screen}
\end{eqnarray}
while in the mixed geometry,
\begin{eqnarray}
E(\tau_{\rm dust})=E_{\rm mix}(\tau_{\rm dust})\equiv
[1-\exp (-\tau_{\rm dust})]/\tau_{\rm dust}
\, .\label{eq:mix}
\end{eqnarray}
Since $E_{\rm mix}>E_{\rm sc}$ for a given $\tau_{\rm dust}$,
the screen geometry shields the UV light more efficiently.
Moreover, the exponential behaviour of $E_{\rm sc}$
results in a strong cut-off of UV light at a certain
time when a significant amount of dust is accumulated,
while $E_{\rm mix}\sim 1/\tau_{\rm dust}$ for
$\tau_{\rm dust}\gg 1$, which means that UV light originating
from the ``surface'' always escapes.
We do not treat a clumpy dust distribution,
since inhomogeneity changes average optical depths
in ways that depend on the specific geometry 
(Natta \& Panagia \cite{natta84}); such a treatment is
beyond the scope of this paper.

$L_{\rm UV,0}$ is assumed to be equal to the total luminosity of
OB stars, whose mass is larger than 3 $M_\odot$ (Cox \cite{cox00}):
\begin{eqnarray}
L_{\rm UV,0}(t)=\int_{3~M_\odot}^{\infty}dm\,\int_0^{\tau_m}
dt'\, L(m)\,\phi (m)\,\psi (t-t')\, ,\label{eq:uv_lum}
\end{eqnarray}
where $L(m)$ is the stellar luminosity as a function of stellar
mass ($m$). This UV luminosity is also used in
Eq.~(\ref{eq:L_UV}) to estimate the UV luminosity $L_{\rm UV}$
after the dust absorption, and the same $L_{\rm UV}$ is also
in Eq.~(\ref{eq:isrf}). For $L(m)$, we adopt the model of
zero-metallicity stars without mass loss in
Schaerer (\cite{schaerer02}). Assuming that all the absorbed
energy is reemitted in the IR, the IR luminosity $L_{\rm IR}$
becomes:
\begin{eqnarray}
L_{\rm IR}=L_{\rm UV,0}-L_{\rm UV}=L_{\rm UV,0}[1-E(\tau_{\rm dust})]
\, .
\end{eqnarray}

\subsubsection{Dust temperature}

Another important quantity representative for the radiative
properties of a star-forming region is dust temperature.
The equilibrium dust temperature is determined from the
balance between incident radiative energy and radiative
cooling. The equilibrium temperature is
expressed as follows
(Takeuchi et al.\ \cite{takeuchi03}):
\begin{eqnarray}
T_{\rm dust}\simeq\left(\frac{hc}{\pi k}\right)\left\{
\frac{63u_{\rm UV}Q_{\rm UV}}{64\pi(2\pi Aa)hc}\right\}^{1/6}\, ,
\label{eq:T_dust_eq}
\end{eqnarray}
where the UV radiation field is calculated by
Eq.~(\ref{eq:isrf}) and the dimensionless dust emissivity in
the IR is assumed to satisfy\footnote{Strictly speaking,
this is valid for $\lambda\ga20\,\mu$m for silicate grains,
and $\lambda\ga50\,\mu$m for carbonaceous ones.}
\begin{eqnarray}
Q_{\rm IR}(a,\,\lambda)=\frac{2\pi Aa}{\lambda^2}\, .
\end{eqnarray}
For silicate grains $A=1.34\times 10^{-3}$ cm (Drapatz \& Michel
\cite{drapatz77}), while for 
carbonaceous grains $A=3.20\times 10^{-3}$ cm (Draine \& Lee
\cite{draine84}; Takeuchi et al.\ \cite{takeuchi03}). The
derivation of Eq.~(\ref{eq:T_dust_eq}) 
is found in the Appendix. The numerical
expression for $T_{\rm dust}$ is
\begin{eqnarray}
T_{\rm dust} & = & 14.6\left(
\frac{u_{\rm UV}}{4.0\times 10^{-14}~{\rm erg~cm}^{-3}}
\right)^{1/6}\nonumber\\
& & \times\left(
\frac{A}{3.2~{\rm cm}\times 10^{-3}}\right)^{-1/6}\left(
\frac{a}{0.03~\mu{\rm m}}\right)^{-1/6}\nonumber\\
& & {}\times Q_{\rm UV}^{1/6}~{\rm K}\, ,
\label{eq:T_dust}
\end{eqnarray}
where we explicitly express the dependence on $a$. 
The equilibrium dust temperature is seen to be rather insensitive
to changes in the input parameters.
We adopt $A=3.2\times 10^{-3}$ cm since
that is the grain radius appropriate for graphites
(Todini \& Ferrara \cite{todini01}). With this value,
we obtain
\begin{eqnarray}
Q_{\rm IR}(a,\,\lambda )=6.03\times 10^{-4}\left(
\frac{a}{0.03~\mu{\rm m}}\right)\left(
\frac{\lambda}{100~\mu{\rm m}}\right)^{-2}\, .
\end{eqnarray}
For a grain size of 0.03~$\mu$m (carbonaceous grains), and 
a density $\delta\,=\,2$\,g\,cm$^{-3}$ (Draine \& Lee \cite{draine84}),
this corresponds to
the following value of dust mass-absorption coefficient,
$\kappa_\nu (\equiv 3Q_{\rm IR}/4a\delta )$
(Hildebrand \cite{hildebrand83}):
\begin{eqnarray}
\kappa_\nu =75\left(
\frac{\lambda}{100~\mu{\rm m}}
\right)^{-2}\left(\frac{\delta}{2~{\rm g}~{\rm cm}^{-3}}
\right)^{-1}~{\rm cm}^2~{\rm g}^{-1}\, .
\end{eqnarray}
Although there is little observational constraint on
$Q_{\rm IR}$ and $\kappa_\nu$ of dust produced by SNe II,
we can compare our $\kappa_\nu$ with the data of Galactic or
extragalactic observations.
At $\lambda =850~\mu$m, $\kappa_\nu\simeq 1.0$ cm$^2$ g$^{-1}$,
while observational data suggest $\kappa_\nu\simeq 0.7$--2.4
cm$^2$ g$^{-1}$ (Alton et al.\ \cite{alton01};
James et al.\ \cite{james02}).
Bianchi et al.\ (\cite{bianchi99}) find that
$Q_{\rm IR}$ around $\lambda =100~\mu$m is roughly
$\sim 10^{-3}$--$5\times 10^{-4}$ if we assume
$Q$ at $V$ band is $\sim 1$ (see also
Fig.\ 5 of James et al.\ \cite{james02} and references
therein).

\subsection{Initial conditions}

For the onset of star formation, the gas must be cool. The
molecular hydrogen cooling eventually cools gas down to
$T\sim 300$ K (e.g., Tegmark et al.\ 1997). During the phase
of efficient molecular formation in gas, the ionisation
fraction $x$ is roughly $\sim 10^{-4}$ and the molecular
fraction is $\fH2\sim 10^{-3}$. After this phase, the ionisation
degree $x$ decreases and the gas phase reaction stops. Since stars
begin to form during molecular formation, we set the initial
values as $x=10^{-4}$, $T=300$ K, and $\fH2 =10^{-3}$. However,
the gas starts to be ionised and the molecules begins to
dissociate soon ($<1$ Myr) after the onset of star formation,
and the results do not depend strongly on the initial conditions 
for $x$ and $\fH2$.
The initial temperature does not change the
evolution of gas as long as $T\la 10^3$ K initially.

\subsection{Selection of parameters}

For the UV light, because of the short wavelength, we can assume
that $Q_{\rm UV}\simeq 1$. For the grain radius, we examine
the following two cases:
small dust ($a=0.03~\mu$m) proposed by
Todini \& Ferrara (\cite{todini01}), and large dust
($a=0.1~\mu$m) possibly produced even in SNe II (Nozawa et al.\
\cite{nozawa03}). We fix $m_{\rm dust}=0.4~M_\odot$
(Section \ref{subsec:dust_formalism}), and adopt
$\delta\simeq 2$ g cm$^{-3}$ (Draine \& Lee \cite{draine84}).
As mentioned in Section \ref{subsec:dust_formalism},
the timescale of dust effects (enhancement of
molecular content, enhancement of UV shielding, etc.)
is roughly inversely proportional to $m_{\rm dust}$.

The free parameters which remain to be determined are
$r_{\rm SF}$, $M_{\rm gas}$, and $\epsilon_{\rm SF}$. Since
our aim is to investigate the properties of ``active'' and
``passive'' star-forming regions, which correspond to compact
and diffuse regions, respectively, we should examine various
$r_{\rm SF}$.
In order to concentrate on the effect of compactness, the gas
mass is fixed. A representative value for a star-forming
region in BCDs can be taken as $M_{\rm gas}=10^7~M_\odot$. 
This is a typical value for gas masses derived for star-forming regions in
\sbs\ and \izw\, representative extremely metal-poor
BCDs (Section \ref{subsec:sbs}).
In order to concentrate on the variation of
$r_{\rm SF}$, we therefore fix $M_{\rm gas}=10^7~M_\odot$, and
examine a region size range typical of BCDs: 
$r_{\rm SF}=30$ pc, 100 pc, and 300 pc, corresponding to
the density $n\sim 3000~{\rm cm}^{-3}$, 100 cm$^{-3}$,
and 3 cm$^{-3}$, respectively, with
$M_{\rm gas}=10^7~M_\odot$. The most compact case is
representative of the ``active'' class, while the most
diffuse case is typical of the ``passive'' class.

The remaining free parameter is the star formation efficiency
$\epsilon_{\rm SF}$, which we assign a value
$\epsilon_{\rm SF}=0.1$ (e.g., Ciardi et al.\ \cite{ciardi00};
Ferrara et al.\ \cite{ferrara00};
Inoue et al.\ \cite{inoue00}; Barkana \cite{barkana02};
Scannapieco et al.\ \cite{evan03};
Salvaterra \& Ferrara \cite{salvaterra03}) unless otherwise
stated. Roughly speaking, the timescale on which the dust
effects appear scales inversely with $\epsilon_{\rm SF}$.
For example, the rapid increase of
molecules and drop of temperature appear on a timescale
proportional to $1/\epsilon_{\rm SF}$. 
It is possible that
$\epsilon_{\rm SF}$ may change significantly as a function of
time, but our simple assumption
of constant $\epsilon_{\rm SF}$ is a useful first-order
approximation.
$\epsilon_{\rm SF}$ and the star formation rate should be
regarded as an average over the timescale in which we
are interested. High and low star formation efficiencies
are investigated in the models of \izw\ and \sbs\ in
Section \ref{subsec:sbs}.

These values are all representative of metal-poor star-forming
regions. Typically, BCDs have 
$\psi\sim 0.01$--1 $M_\odot~{\rm yr}^{-1}$,
$r\sim 0.1$--3 kpc and $n\sim 10$--1000 cm$^{-3}$
(Popescu et al.\ \cite{popescu99};
Hopkins et al.\ \cite{hopkins02}). 

\section{Results}\label{sec:result}

Here we illustrate the results for the time evolution of star-forming
regions. First we adopt $a=0.03~\mu$m and the screen dust
distribution. 
Then, in Section \ref{subsec:largegrain},
we investigate the dependence on grain size by assuming larger grains
($a=0.1~\mu$m) and on geometry by examining the 
effects of a mixed dust distribution (Eq. \ref{eq:mix}). 

For continuous star-forming activity, the gas should continue to
collapse. With a fixed density, the
typical mass for the collapse, i.e., the Jeans mass, is
determined by the gas temperature. Therefore, the evolution
of the gas temperature is one of the most important processes
for understanding the fate of a star-forming region. In the
following, we show the time evolution of gas temperature ($T$)
and molecular fraction ($\fH2$) for various region radii.
Because we have assumed fixed values for $M_{\rm gas}$,
the density $n_{\rm H}$ is thus determined by setting $r_{\rm SF}$.
Moreover, since we have also fixed 
$\epsilon_{\rm SF}$, 
the resulting star-formation rate $\psi$, given by
Eq.~(\ref{eq:sfr}), also depends on $r_{\rm SF}$ through the
density $n$ dependence.

\begin{figure*}
\includegraphics[width=6cm]{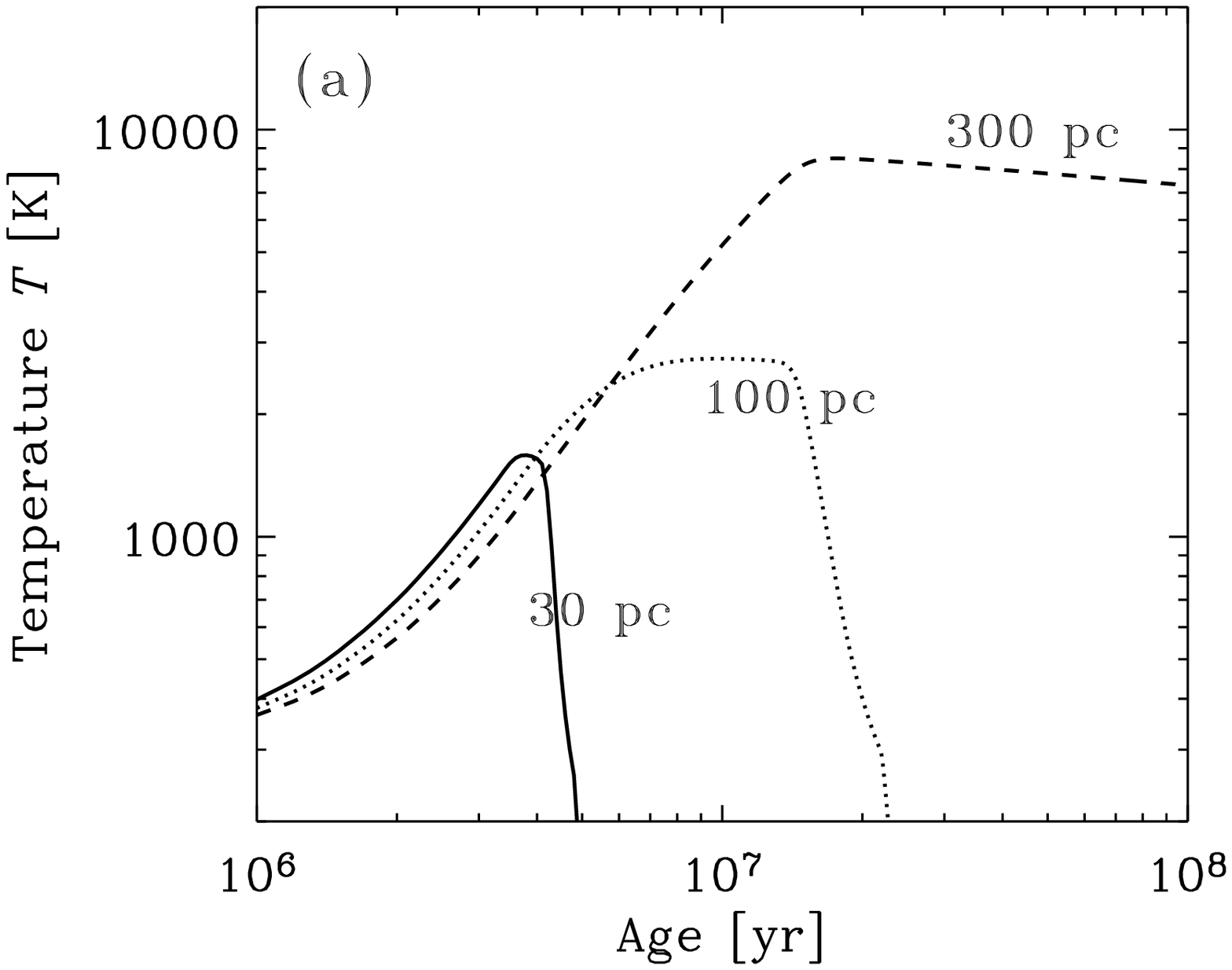}
\includegraphics[width=6cm]{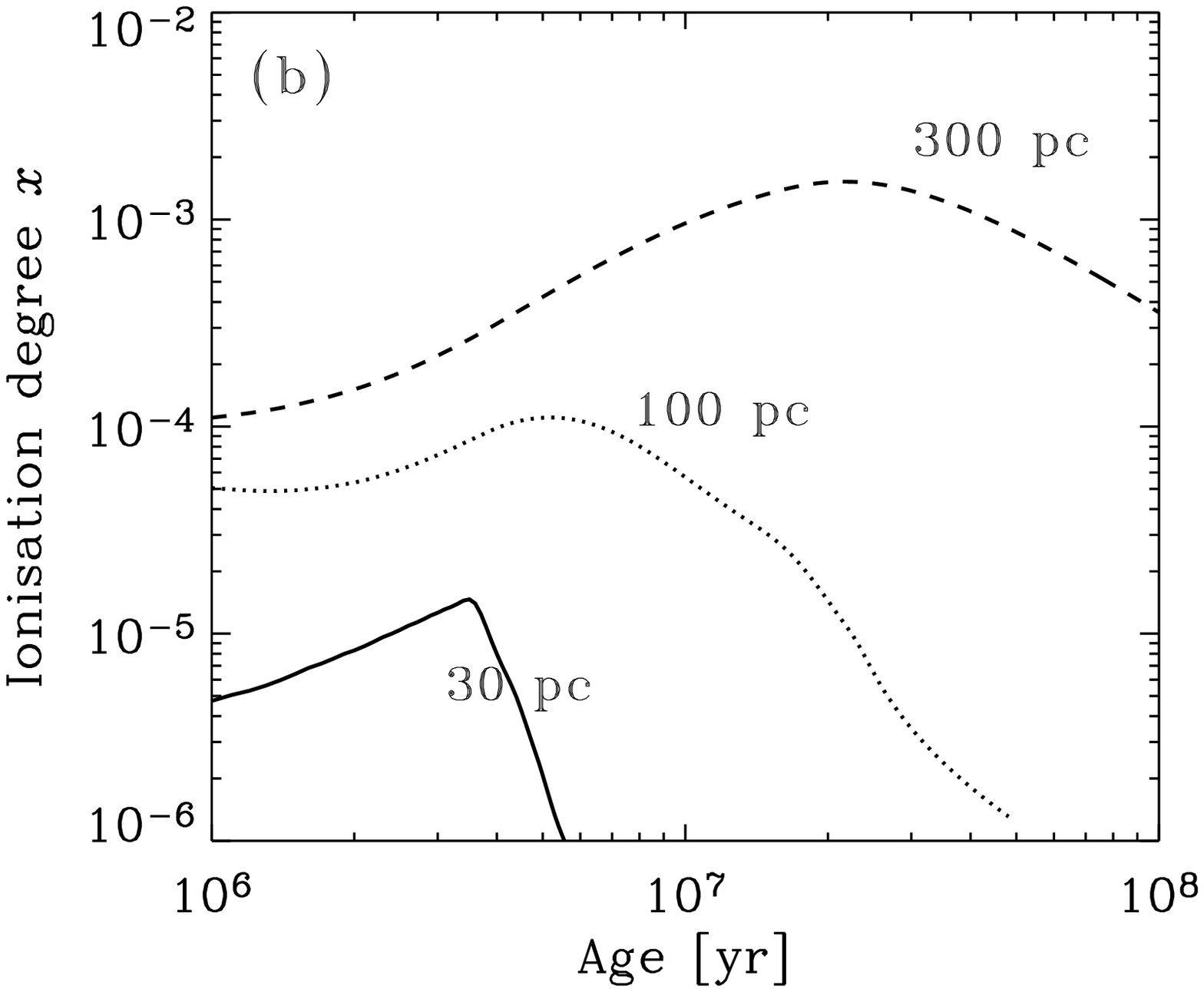}
\includegraphics[width=6cm]{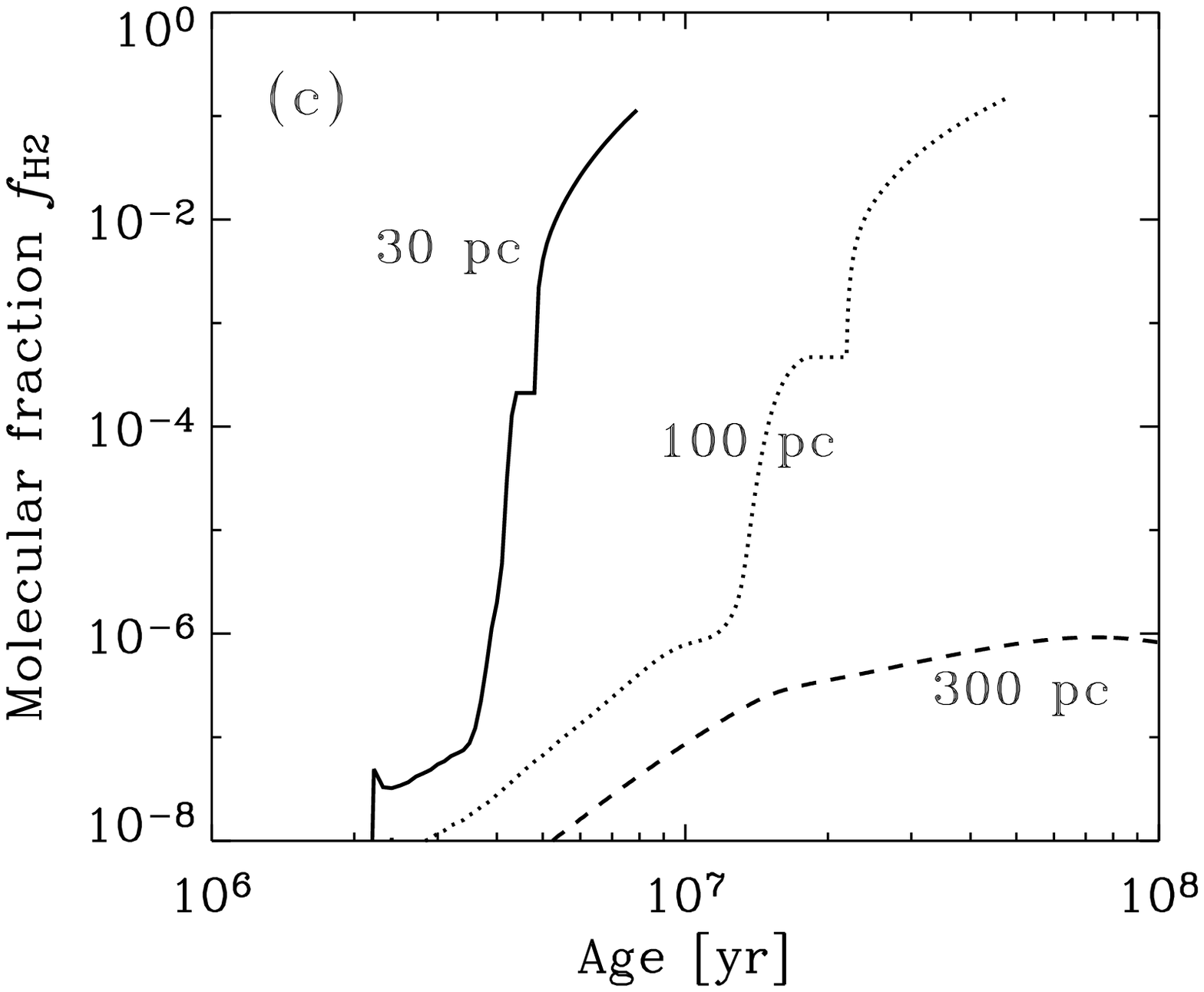}
\caption{Time evolution of {\bf a)} gas temperature, {\bf b)}
ionisation degree, and {\bf c)} molecular fraction for various
radii of the star-forming region (solid, dotted, and dashed
lines for $r_{\rm SF}=30$ pc, 100 pc, and 300 pc, respectively).
The gas mass and the star formation efficiencies
are assumed to be $1.0\times 10^7~M_\odot$ and
0.1, respectively.}
\label{fig:radius_dep}
\end{figure*}

\subsection{Gas state}\label{subsec:state}

In Fig.~\ref{fig:radius_dep}, we show the evolution of (a)
gas temperature, (b) ionisation degree, and (c) molecular
fraction for $r_{\rm SF}=30$, 100, 300 pc (solid, dotted, and
dashed lines, respectively). We stop our calculation at
$t=t_{\rm gas}/2$, when our assumption that $M_{\rm gas}$
is constant becomes invalid (see Eq. \ref{eq:tgas}). 
Therefore the lines are truncated at
8 Myr and 50 Myr for $r_{\rm SF}=30$ pc and 100 pc, respectively.
We observe that as the region size
becomes smaller (i) the gas remains cooler, (ii) the ionisation
degree remains lower, and (iii) the molecular fraction
increases more rapidly. 

In the time evolution of these three parameters, accumulation
of dust grains plays a fundamental role. The dust accumulation
in a compact region results in a large optical depth and a
large density of dust grains. As a result, 
the molecular formation tends to increase
rapidly, because of the large density of dust grains and the
consequent large shielding of dissociating photons. Since the UV photons are
efficiently shielded by dust grains, the gas is kept cool and
the ionisation degree remains low.

The temporal behaviour of the molecular fraction is explained
as follows (see Fig.\ \ref{fig:radius_dep}c). The first
drastic increase of the molecular fraction, seen around 4 and
13 Myr for $r_{\rm SF}=30$ and 100 pc, respectively, is due to
the activation of the gas-phase reactions caused by
the increase of ionisation degree and the temperature drop. 
The molecular formation is suppressed for $r_{\rm SF}=300$ pc
because the gas temperature remains high $\sim\,10^4$ K.
The second drastic
increase of the molecular fraction, seen around 5 and 20 Myr
for $r_{\rm SF}=30$ and 100 pc, respectively
(for $r_{\rm SF}=300$ pc, the second increase is not
before 100 Myr), is due to the onset of molecule
formation on dust grains: dust grains shield the ionising
photons and facilitate the gas cooling down to 300 K. At this
temperature, the \H2 formation on dust grains becomes possible. For
$r_{\rm SF}=300$ pc, representative of the
passive mode, \H2 formation occurs only in the gas phase
with a low reaction rate because of the high gas temperature.

\subsection{Luminosities}

\begin{figure*}
\includegraphics[width=9cm]{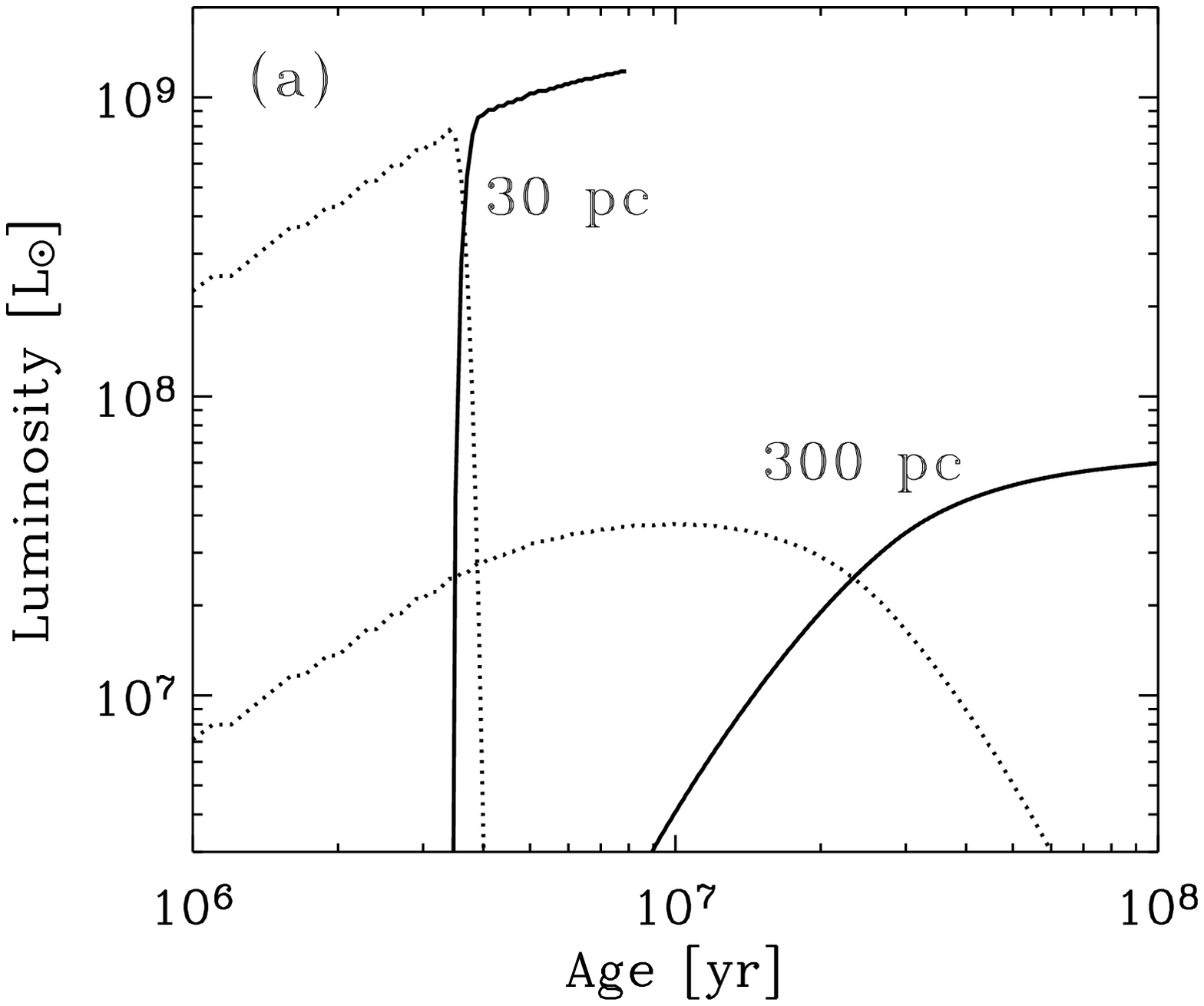}
\includegraphics[width=9cm]{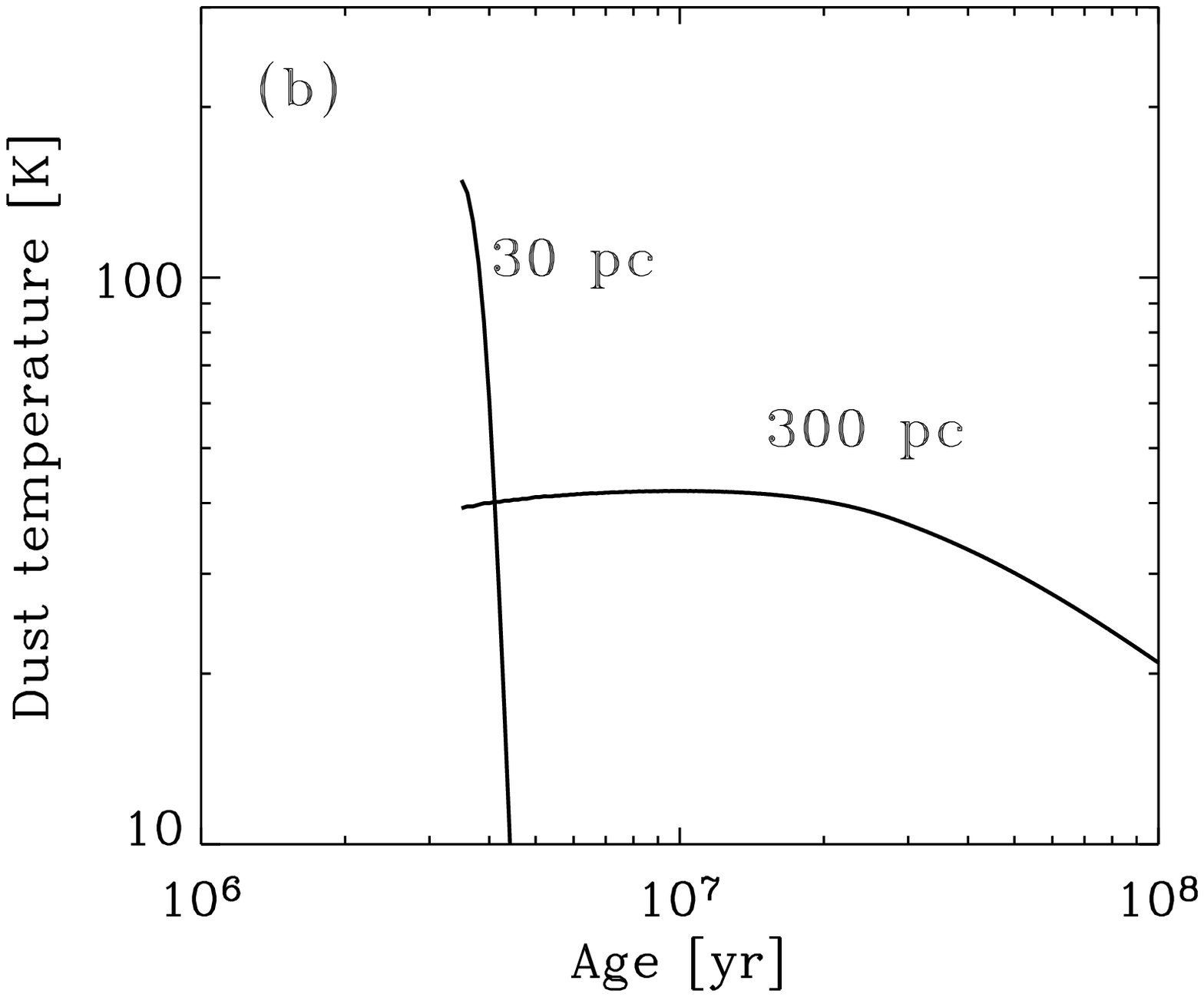}
\caption{{\bf a)} Time evolution of IR and UV luminosities
(solid and dotted lines, respectively).
The left lines are
for the compact ($r_{\rm SF}=30$ pc) case and the right lines
are for diffuse ($r_{\rm SF}=300$ pc) case. The gas mass is
assumed to be $10^7~M_\odot$ (same as
Fig.\ \ref{fig:radius_dep}). {\bf b)} Time evolution of dust
temperature. The upper and lower lines are for
$r_{\rm SF}=30$ pc and $r_{\rm SF}=300$ pc, respectively.
The lines start from the age of around 3 Myr, when
dust begins to be produced. The screen dust distribution and
the small dust grain ($a=0.03~\mu$m) are assumed for both
panels.}
\label{fig:luminosity}
\end{figure*}

To see how much UV light is shielded by dust, we show
the evolution of IR and UV luminosities in
Fig.~\ref{fig:luminosity}a for $r_{\rm SF}=30$ pc and
$r_{\rm SF}=300$ pc, corresponding to the active and passive
modes, respectively. We see that the IR vs.\ UV luminosity
ratio increases more rapidly in the
compact ($r_{\rm SF}=30$ pc) region than in the diffuse
($r_{\rm SF}=300$ pc) one. This is due to the large
dust optical depth in compact star-forming regions. The
efficient shielding of UV in compact regions also keeps the
molecular fraction high since UV dissociation is suppressed.

The evolution of dust temperature determined from
Eq.~(\ref{eq:T_dust}) is also shown in
Fig.~\ref{fig:luminosity}b for $r_{\rm SF}=30$ pc
and 300 pc. The lines start from the onset of the first
SNe II (i.e., $t=3$ Myr), when dust production begins. The
dust temperature is initially $\sim 100$ K and decreases
rapidly for the compact
``active'' case. This rapid decrease is due to the
strong shielding of UV light by dust grains. The dust
temperature remains $\sim 50$ K for the ``passive'' case.

\subsection{Chemical enrichment}\label{subsec:chemical}

\begin{figure*}
\includegraphics[width=9cm]{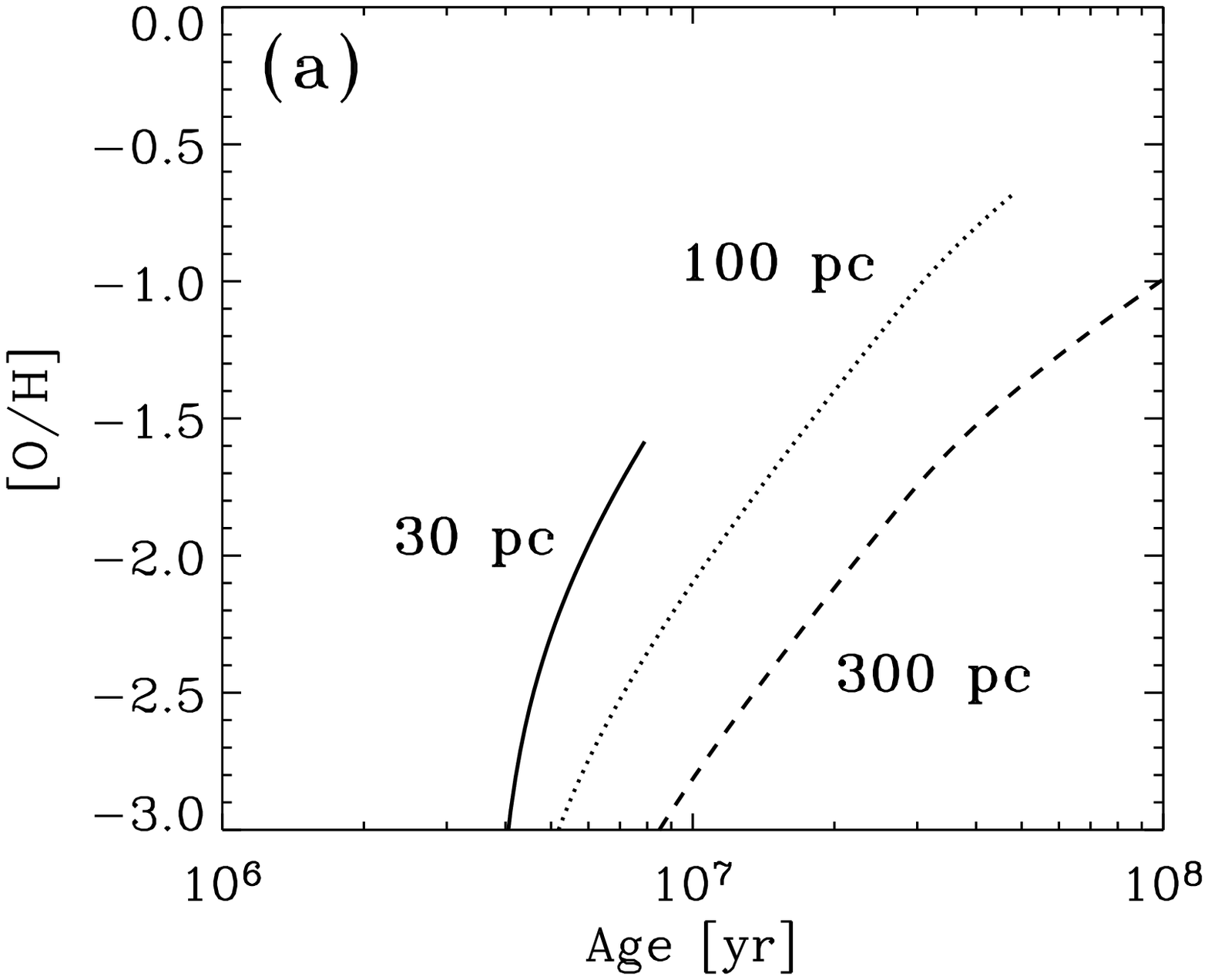}
\includegraphics[width=9cm]{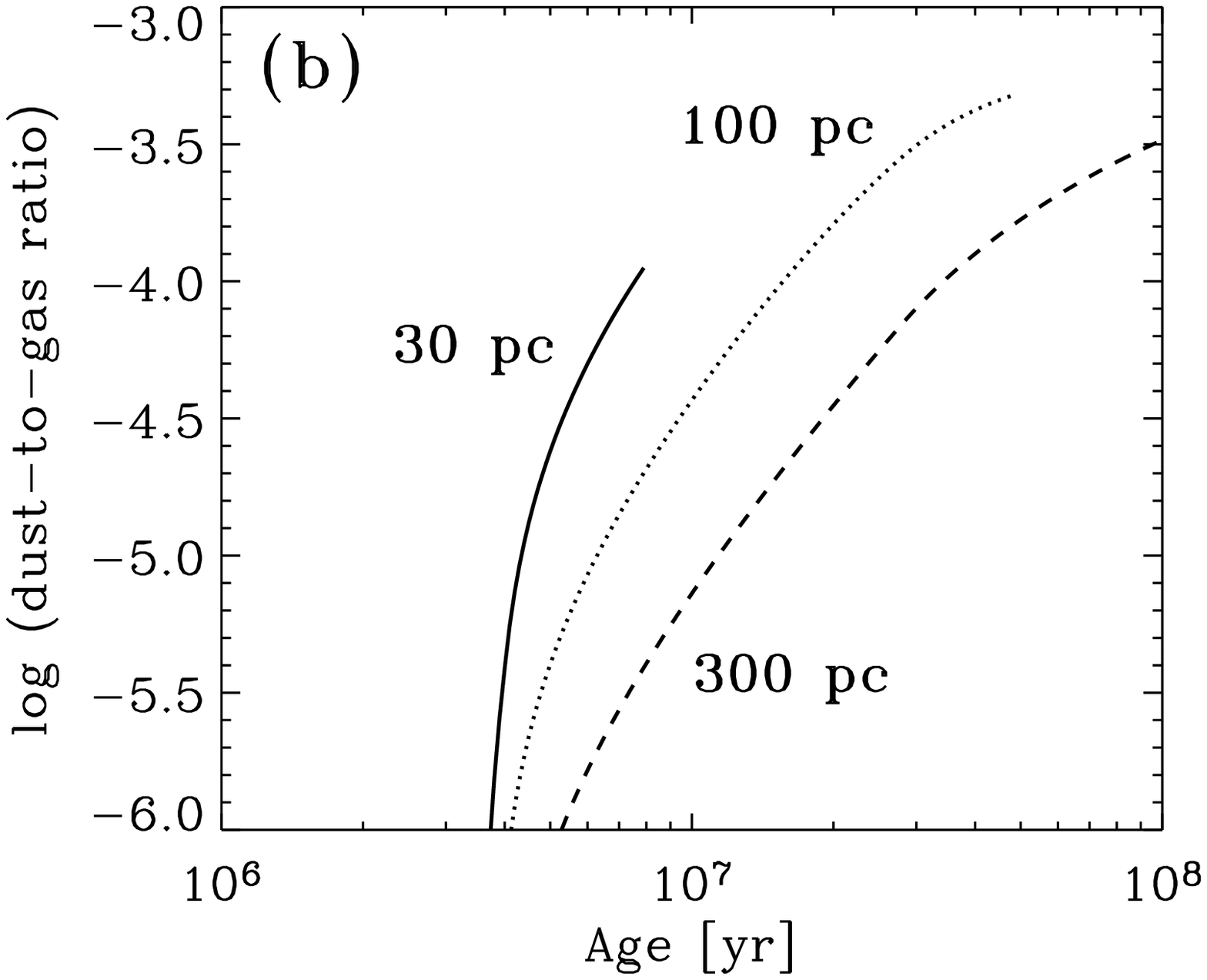}
\caption{Time evolution of oxygen abundance in gas phase.
[O/H] is the oxygen abundance in logarithmic scaling,
where ${\rm [O/H]}=0$ is the solar oxygen abundance.
The solid, dotted, and dashed lines are for
$r_{\rm SF}=30$ pc, 100 pc, and 300 pc, respectively.
The gas mass and the star formation efficiency are
assumed to be $10^7~M_\odot$ and 0.1, respectively (same as
Fig.~\ref{fig:radius_dep}).}
\label{fig:metaldust}
\end{figure*}

In order to ascertain if our chemical enrichment model is
capable of explaining the observed metallicities, we
calculate the evolution of oxygen abundance. Since the
optical oxygen emission lines from ionised regions are
often used as a tracer of chemical abundance (e.g.,
Izotov \& Thuan \cite{izotov99}), we calculate the
evolution of oxygen mass (see Section \ref{subsec:metal});
the solar abundance is assumed to correspond to 
$M_{\rm O}/M_{\rm gas}=0.01$ (Cox \cite{cox00}), 
where $M_{\rm gas}$ is assumed to be constant in time.
Fig.~\ref{fig:metaldust}a shows the evolution
of [O/H] for the three region radii depicted in
Fig.~\ref{fig:radius_dep}. The metallicity of the 30-pc region
reaches 1/50 solar in 7 Myr. Not surprisingly,
the metal enrichment in diffuse regions proceeds more slowly:
in particular, the 300-pc region is enriched to 1/50 solar
in $\sim 30$ Myr.

The evolution of dust-to-gas ratio
(${\cal D}\equiv M_{\rm dust}/M_{\rm g}$) is shown in
Fig.~\ref{fig:metaldust}b. We have assumed that the
star-forming regions contain no dust and metals initially. This
assumption holds if a large part of the present dust and
metals is supplied during the present episode of
star formation. This assumption is adopted based on the
picture that BCDs are relatively young galaxies whose major
star formation episodes have occurred quite recently ($\la 100$ Myr;
e.g., Tomita et al.\ \cite{tomita02};
Takeuchi et al.\ \cite{takeuchi04}). This point will be further
discussed when we compare our model with the observational
properties of BCDs (Section \ref{subsec:sbs}).

\subsection{Grain size and distribution geometry \label{subsec:largegrain}}

Nozawa et al.\ (\cite{nozawa03}) have recently
pointed out that grains larger than predicted by Todini \&
Ferrara (\cite{todini01}) can form even in SNe II.
Grains may be as large as $a\sim 0.1~\mu$m, and therefore,
we also examine this case. 

\begin{figure*}
\includegraphics[width=9cm]{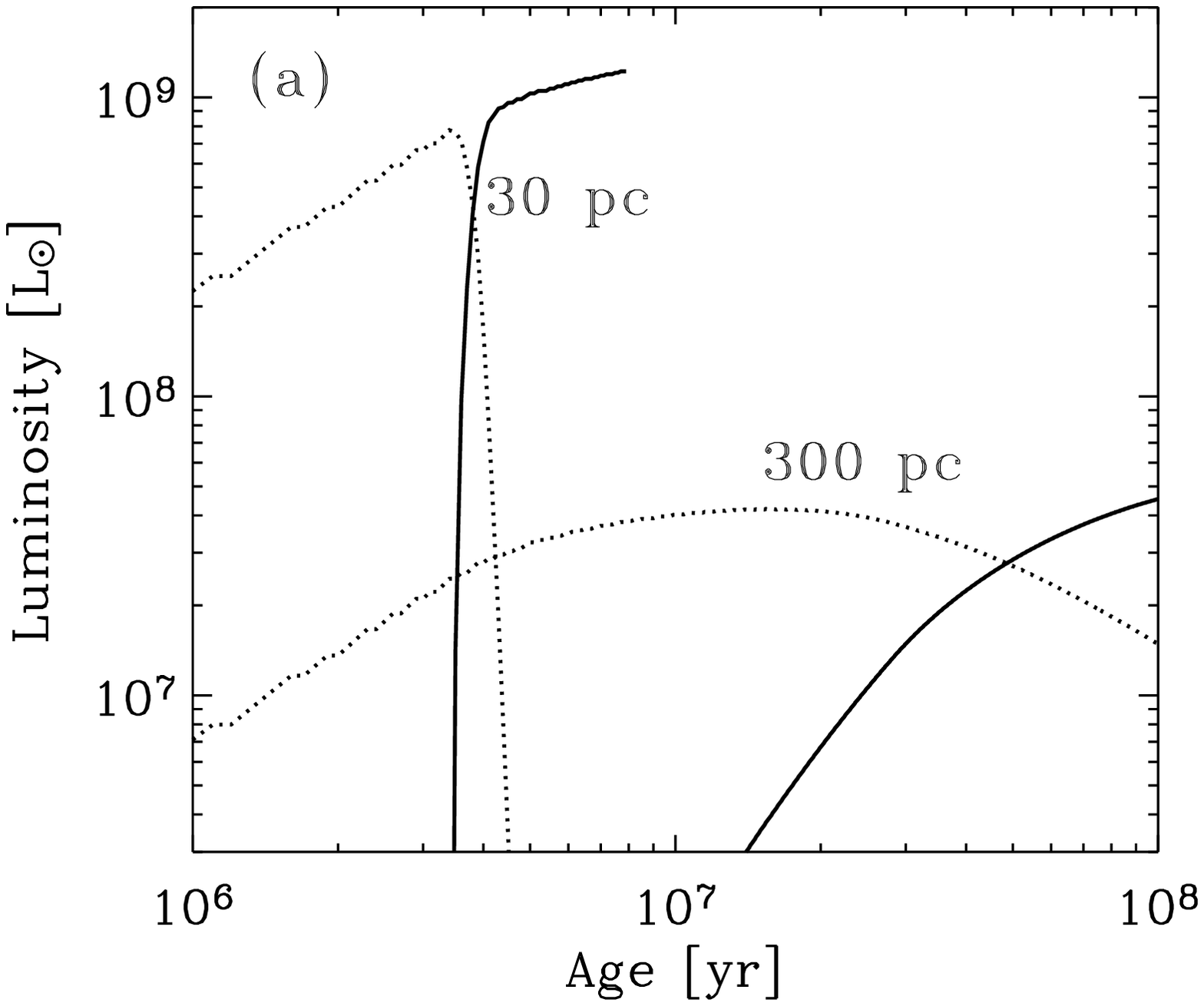}
\includegraphics[width=9cm]{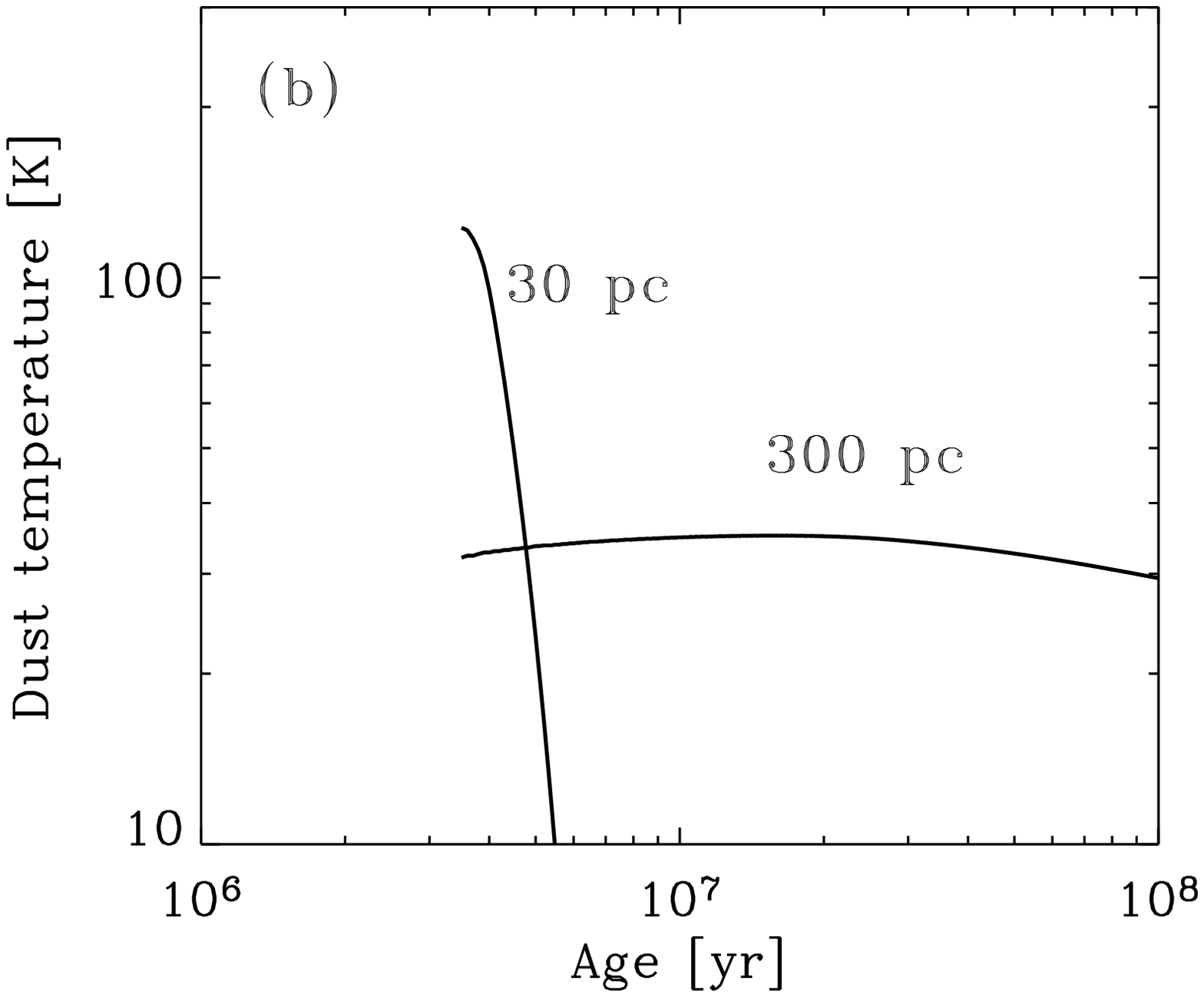}
\caption{{\bf a)} Same as Fig.~\ref{fig:luminosity}a, but for 
large grains ($a=0.1~\mu$m). {\bf b)} Same as
Fig.~\ref{fig:luminosity}b, but for large grains.}
\label{fig:lumi_large}
\end{figure*}

In order to evaluate the effect of shielding, we show the evolution
of UV and IR luminosities for $a=0.1~\mu$m with the
screen geometry in Fig.~\ref{fig:lumi_large}a. With a
fixed dust mass, the dust optical depth against the UV photons,
$\tau_{\rm dust}$, is
proportional to $a^{-1}$, because $n_{\rm dust}\propto a^{-3}$
and $\tau_{\rm dust}\propto a^2 n_{\rm dust}$ (Eq.\ \ref{eq:def_tau0}).
Therefore, if the grain size is larger, more
UV radiation escapes without being absorbed by
dust grains. Consequently, the IR luminosity becomes dominant
relative to $L_{\rm UV}$
later for $a=0.1~\mu$m than for $a=0.03~\mu$m, although the
difference is not significant especially for the compact
($r_{\rm SF}=30$ pc) region. For $r_{\rm SF}=30$ pc,
the IR luminosity dominates as soon as the dust
production begins ($\sim 3$--4 Myr).

We also show the evolution of dust temperature in
Fig.~\ref{fig:lumi_large}b. The dust temperature drops rapidly
in compact regions ($r_{\rm SF}=30$ pc) because of the efficient
shielding of UV radiation. 
For diffuse regions ($r_{\rm SF}=300$ pc), the temperature drop 
relative to small grains is not
significant because the shielding is inefficient.
Owing to the $a^{-1/6}$ dependence of dust temperature
(Eq.\ \ref{eq:T_dust}), the dust temperature is lower
for $a=0.1~\mu$m than $a=0.03~\mu$m.

\begin{figure*}
\includegraphics[width=9cm]{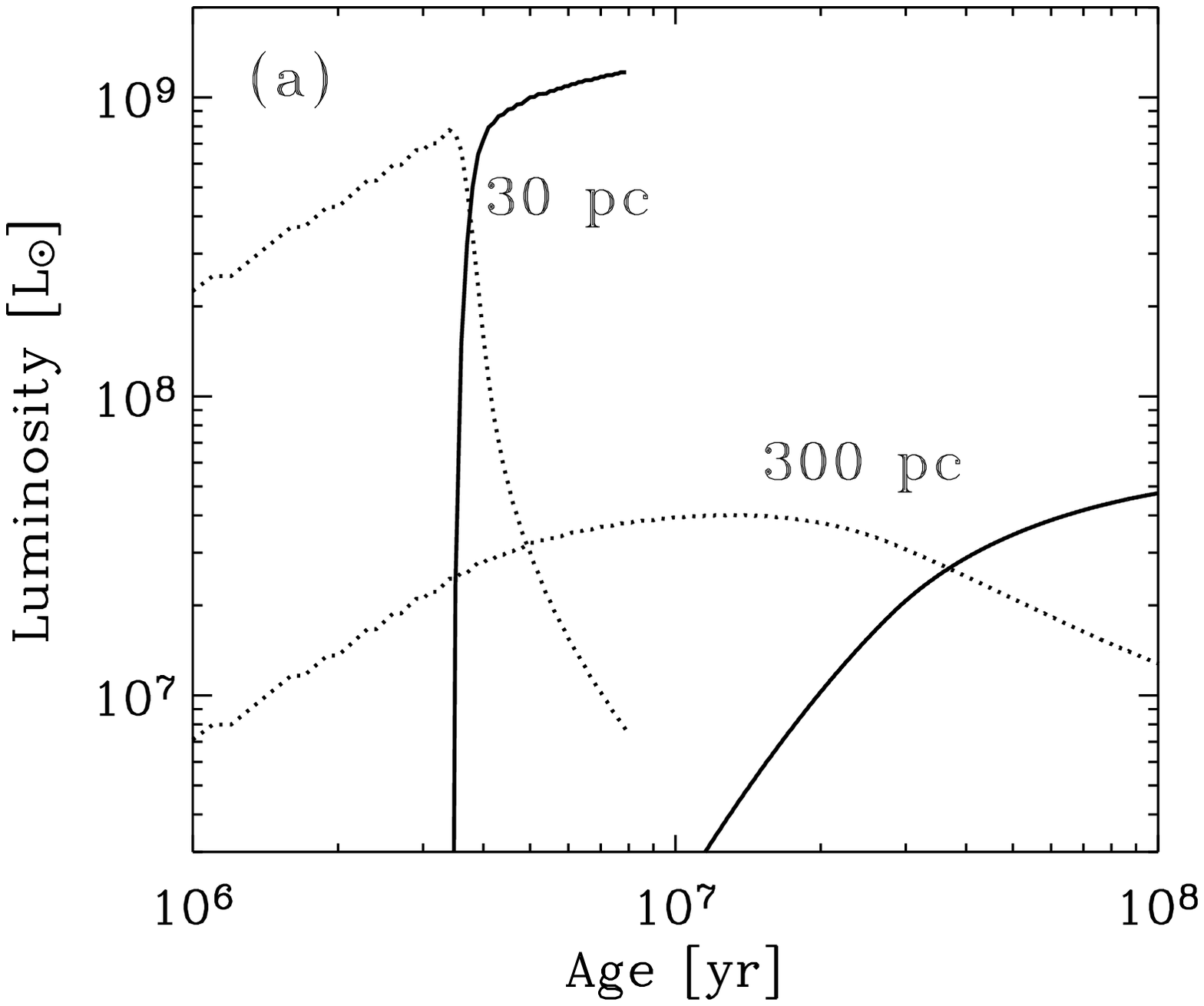}
\includegraphics[width=9cm]{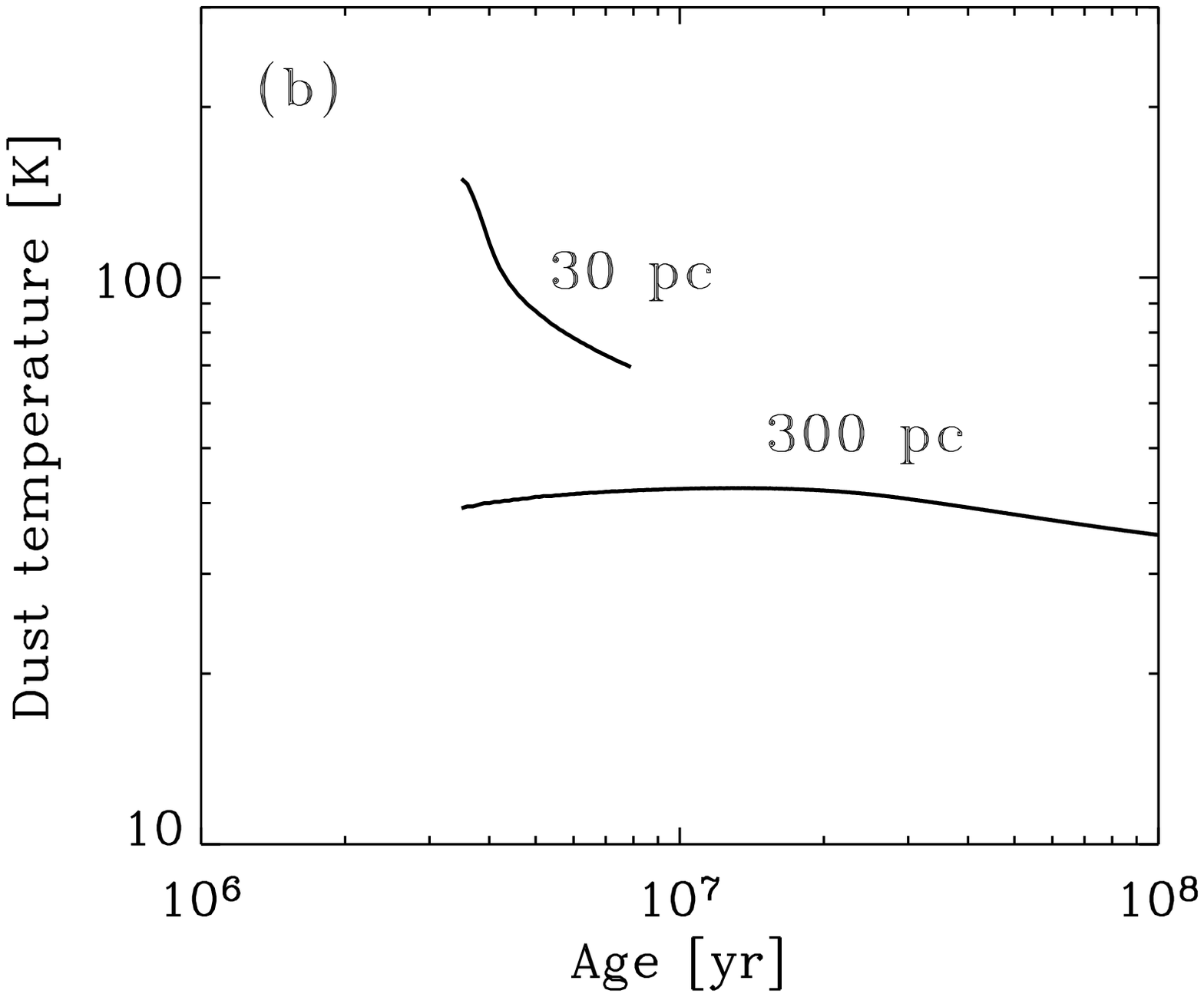}
\caption{{\bf a)} Same as Fig.~\ref{fig:luminosity}a, but for
the mixed geometry of dust distribution. {\bf b)} Same as
Fig.~\ref{fig:luminosity}b, but for the mixed geometry of dust
distribution.}
\label{fig:lumi_mix}
\end{figure*}

Next, in order to investigate the effect of dust distribution
geometry, we examine the mixed geometry using the attenuation
function expressed in Eq.~(\ref{eq:mix}) rather than
Eq.~(\ref{eq:screen}).
The evolution of UV and IR luminosities is shown in
Fig.~\ref{fig:lumi_mix}a. The UV light decreases more mildly
in the mixed geometry than with the
screen geometry, because UV light originating from the
``surface'' of star-forming regions always escapes as
described in Section \ref{subsubsec:UV_IR}.
The evolution of dust temperature is shown in
Fig.~\ref{fig:lumi_mix}b. We see that the strong
``exponential'' drop of dust temperature with the screen geometry 
(Figs.\ \ref{fig:luminosity}b and \ref{fig:lumi_large}b)
is not seen in the mixed case because of the milder shielding of 
UV light. 

\begin{figure*}
\includegraphics[width=6cm]{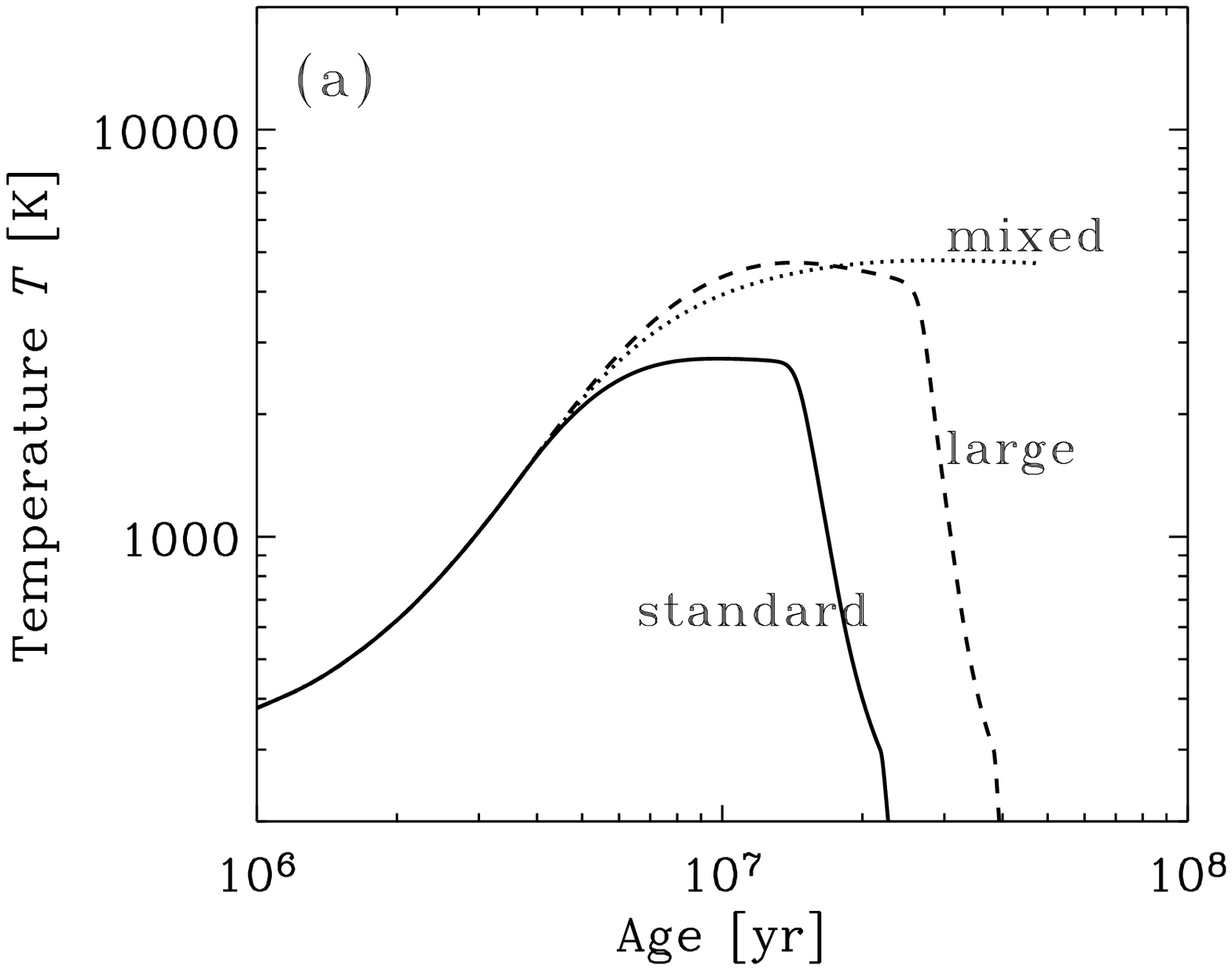}
\includegraphics[width=6cm]{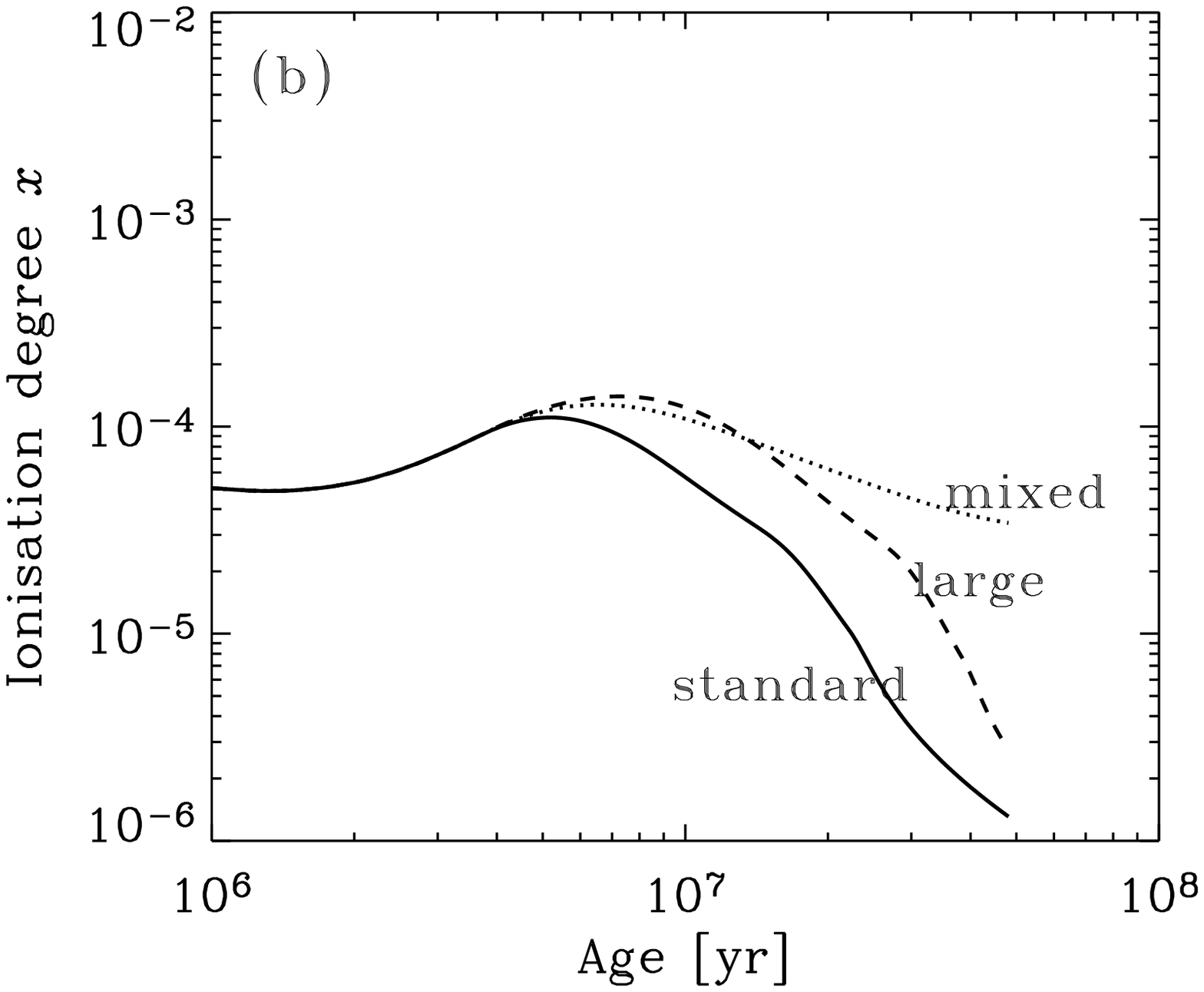}
\includegraphics[width=6cm]{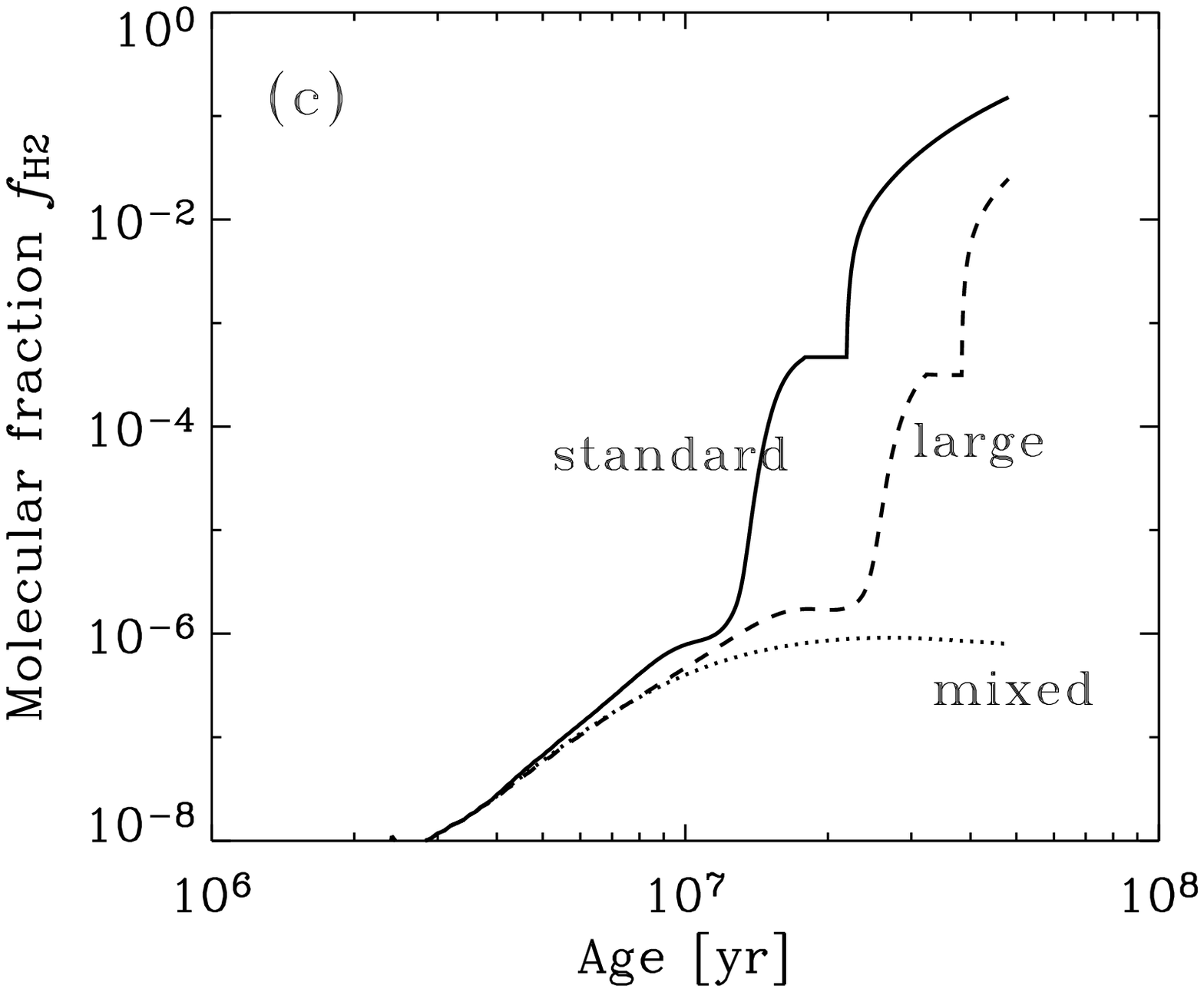}
\caption{Time evolution of {\bf a)} gas temperature,
{\bf b)} ionisation degree, and {\bf c)} molecular fraction
for $r_{\rm SF}=100$ pc and $M_{\rm gas}=10^7~M_\odot$. The
solid, dotted, and dashed lines represents the ``standard''
($a=0.03~\mu$m and
screen geometry) case, the mixed dust geometry
($a=0.03~\mu$m), and the large dust grains
of the star-forming region (solid, dotted, and dashed lines
for $r_{\rm SF}=30$ pc, 100 pc, and 300 pc, respectively).
The gas mass and the star formation efficiencies
are assumed to be $1.0\times 10^7~M_\odot$ and
0.1, respectively.}
\label{fig:gas_various}
\end{figure*}

The physical state of the gas is also affected by the assumptions
about dust grains. In order to clarify this point, we compare the
evolution of gas temperature, ionisation degree, and molecular
fraction in Fig.~\ref{fig:gas_various} for the case of
$r_{\rm SF}=100$ pc. The solid lines,
which are the same as the dotted lines in
Fig.~\ref{fig:radius_dep}, show
the case of $a=0.03~\mu$m and the screen geometry
(called ``standard''). The dotted and dashed lines represent
the cases of the mixed geometry (with $a=0.03~\mu$m) and
$a=0.1~\mu$m (with the screen geometry). Because the
UV optical depth of dust grains becomes smaller as the
grain size increases, the heating and dissociation rates
are larger for larger grains, which explains the
behaviour of the dashed lines. For the molecular fraction,
the formation rate on dust grains (Eq.\ \ref{eq:reaction_h2})
decreases because of the reduced dust surface, which also
contributes to the slow increase of $\fH2$ for
the larger grains. In the mixed geometry,
the shielding of UV light is inefficient; the
gas is heated and molecules are dissociated.
Therefore, we conclude that the grain size and distribution
are important to determine the physical state of gas.

To further investigate the effect of dust size and
spatial distribution on our model results, we are planning to 
extend our study to include infrared SEDs. The study of
SEDs has the advantage of being able to treat the
multi-temperature, multi-size grain distributions, as well as 
stochastically heated grains (Takeuchi et al.\ \cite{takeuchi03};
see also Galliano et al.\ \cite{galliano03}). Not only
extragalactic studies but also Galactic studies will help
us understand why \H2 formation is enhanced in actively
star-forming regions (e.g., Habart et al.\ \cite{habart03}).

\section{Active and passive BCDs}\label{sec:bcds}

\subsection{Physical properties}

In the above, we have shown that both compact and diffuse
star-forming regions can keep the ISM cool and rich in molecules.
A compact and dense (typically $n\ga 10^3$ cm$^{-3}$,
$r_{\rm SF}\la 50$ pc, and
$\psi\ga 0.1~M_\odot~{\rm yr}^{-1}$) region forms stars on a
timescale of $\la 10$ Myr. This regime is termed an ``active''
mode in HHTIV. Since grains supplied by SNe II efficiently
shield UV photons, the gas cools and molecule formation is
enhanced. This implies that in such an active region, gas continues
to collapse and stars form in a ``run-away'' mode. On the contrary,
if a star-forming region is diffuse ($n\la 50$ cm$^{-3}$ and
$r_{\rm SF}\ga 100$ pc), the star formation takes place
``quiescently'' ($\psi\la 0.07~M_\odot~{\rm yr}^{-1}$) on a
timescale of $\ga 10^8$ yr. This mode is called a ``passive''
mode in HHTIV. 
The latter regime
inhibits an ``active'' mode for the following two
reasons: first, the gas collapse occurs slowly because
of a long dynamical timescale; second, because of the inefficient
shielding, it is subject to
strong UV heating and photodissociation of molecules resulting from
the star formation activity.

Let us consider our results from the viewpoint of the gas state. In an
active region, independently of the initial conditions,
the efficient dust accumulation lets the gas cool to $T\la 300$ K  
through molecular cooling and UV shielding. Such a
low temperature is crucial for the efficient formation of \H2
(Cazaux \& Tielens \cite{cazaux02}),  
causing a rapid increase of $\fH2$. 
On the contrary, the evolution of a passive region
proceeds without molecular cooling and UV shielding, and remains
hotter ($T\ga5000$\ K). This implies that the gravitational
collapse in diffuse regions is inefficient and
{\it self-regulated}, because
of the stellar heating, especially the photo-ionisation
heating (Lin \& Murray \cite{lin92}).

What determines the size and density of a star-forming region?
With a constant density, the typical length of a
gravitationally bound region is given by the Jeans length
$\lambda_{\rm J}$:
\begin{eqnarray}
\lambda_{\rm J}=100\left(\frac{T}{10^4~{\rm K}}\right)^{1/2}
\left(\frac{n}{100~{\rm cm}^{-3}}\right)^{-1/2}~{\rm pc}\, .
\label{eq:jeans}
\end{eqnarray}
If a star-forming region is photo-heated, the gas
temperature becomes $T\sim 10^4$ K. Therefore, the typical
size of a self-regulated star-forming region is determined
by the gas density as
$\sim 100(n/100~{\rm cm}^{-3})^{-1/2}$, consistent with the
size of passive star-forming regions. Since an active
star-forming region further cools through dust shielding,
the gas collapses. We suggest
that this collapse finally produces the high surface
brightness and compact size observed for active star-forming regions.

Therefore, the initial gas density, which determines the
self-gravitating length (Eq.\ \ref{eq:jeans}) of a star-forming
region, is important for the bifurcation into active and
passive regimes.
However, ambient pressure can also influence the fate of a
star-forming region
(e.g., Elmegreen \& Efremov (\cite{elmegreen97}; 
Elmegreen \& Elmegreen \cite{elmegreen97}
Elmegreen \& Hunter \cite{eh00}) . 
Elmegreen \& Efremov (\cite{elmegreen97}) stress that the 
star formation efficiency is affected by ambient pressure. 
Shock compression may be important to create high-pressure
environments (Elmegreen \& Elmegreen \cite{elmegreen97}).
A high pressure environment favours
the formation of
SSCs (Bekki \& Couch \cite{bekky01}; Billett et al. \cite{billett02}),
which are observationally known
to be associated with strong starbursts (e.g.,
Hunter et al.\ \cite{hunter94}). If a
star-forming region is confined by a high environmental pressure,
the compression of gas can lead to a high density and
a short free-fall time. This means that high pressure
regions tend to engender an active mode of star formation.
Since there is no reason why the ambient pressure 
should be related
to metallicity, this naturally explains why
metallicity is not a primary factor of distinguishing active and
passive modes.

To link ambient pressure and region size, it is appropriate
to utilise the Ebert-Bonnor formalism for self-gravitating,
pressure-bounded isothermal spheres (Ebert \cite{ebert55};
Bonnor \cite{bonnor56}). The relation indicates that if $r_{\rm SF}$
is smaller than the critical radius,
\begin{eqnarray}
r_{\rm c} & = & 0.49\frac{kT}{m_{\rm H}}\left(
\frac{1}{Gp_{\rm ext}}\right)^{1/2}\nonumber\\
& = & 43\left(\frac{T}{10^4~{\rm K}}\right)
\left(\frac{nT}{10^6~{\rm cm}^{-3}~{\rm K}}\right)^{-1/2}\,
~{\rm pc},
\end{eqnarray}
the region becomes unstable and collapses. Therefore, if the
region is compressed roughly down to $2r_{\rm c}\sim 100$ pc,
it becomes unstable, and can
evolve into an active star-forming region.
Another important role of ambient pressure is to avoid the
expansion of the star-forming region and to maintain
a high-density environment. In summary, ambient pressure
has two effects: one is to make the star-forming region
gravitationally unstable, and the other is to keep the density
high which leads to short free-fall times (e.g.,
Elmegreen \cite{elmegreen00}).

\subsection{Two prototypes}\label{subsec:sbs}

One of our main motivations is to clarify the reason why
\sbs\ and \izw\ show different modes of star formation:
active and passive (HHTIV). First, we
``simulate'' these two galaxies with our models.
The necessary quantities for our model are $M_{\rm gas}$
and $r_{\rm SF}$.
The H\,{\sc i} gas mass has been observationally derived as
$\sim 10^9M_\odot$ (Pustilnik et al.\ \cite{pustilnik01})
and $2.6\times10^7~M_\odot$
(van Zee et al.\ \cite{vanzee98})
for \sbs\ and \izw, respectively. However, we should
consider these values as upper limits for $M_{\rm gas}$,
since $M_{\rm gas}$ is the gas mass in the star-forming
regions, not in the entire system. It is difficult to
resolve the star-forming regions in BCDs because they
are generally small.
On the other hand, typical
gas densities in star-forming regions have been derived
from the argument of collisional excitation based on
high (spatial) resolution spectroscopy 
(e.g., Izotov et al. \cite{izotovetal99}).
Although this argument is biased toward the ionised region
close to massive stars, the gas density derived in this
method can be considered to be representative of the 
entire star-forming region.
Therefore, we will attempt to distinguish active and passive star 
formation and constrain the gas mass $M_{\rm gas}$
on the basis of observed region sizes $r_{\rm SF}$
(e.g., HHTIV) and gas densities.

\subsubsection{\izw}

\begin{figure*}
\includegraphics[width=6cm]{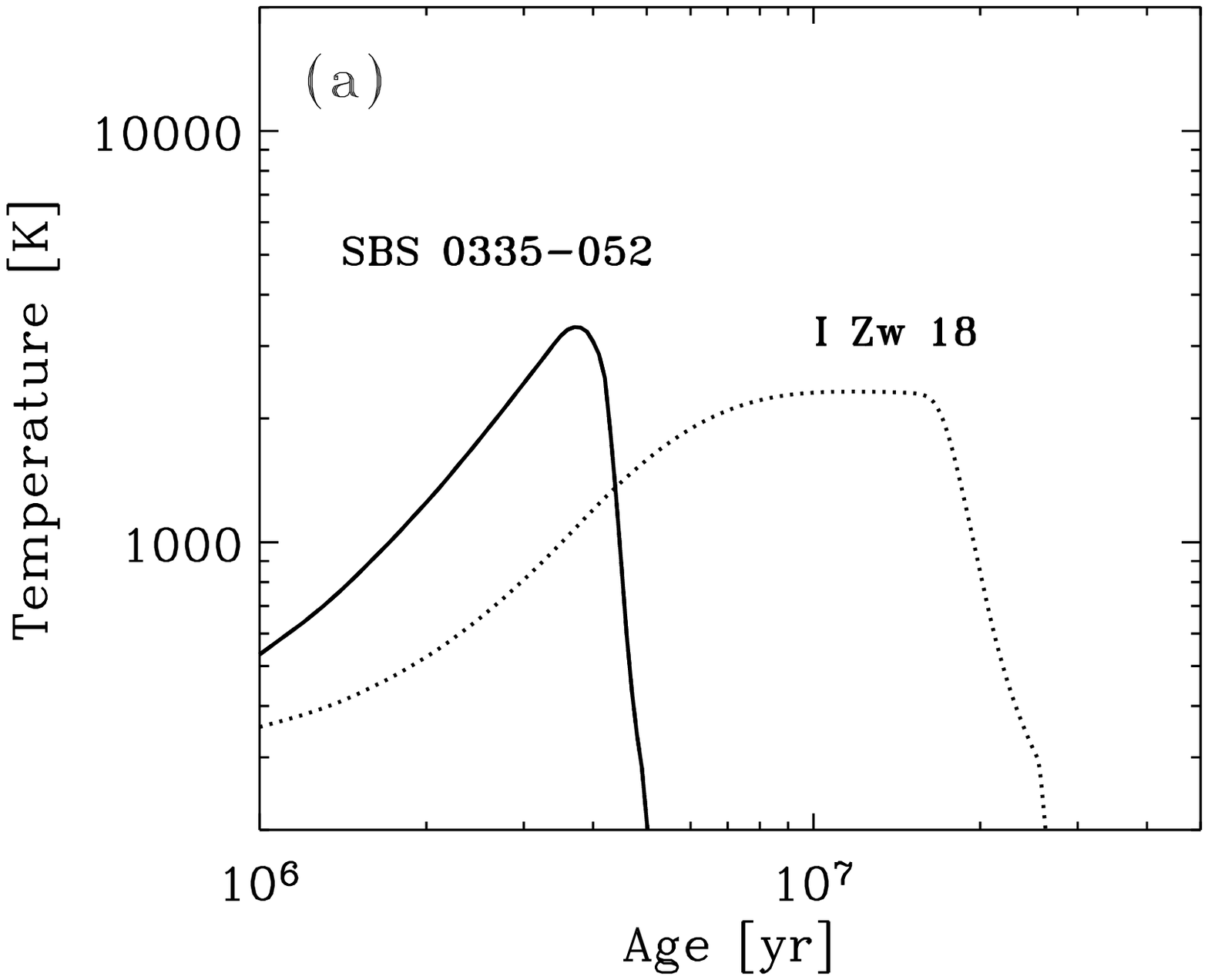}
\includegraphics[width=6cm]{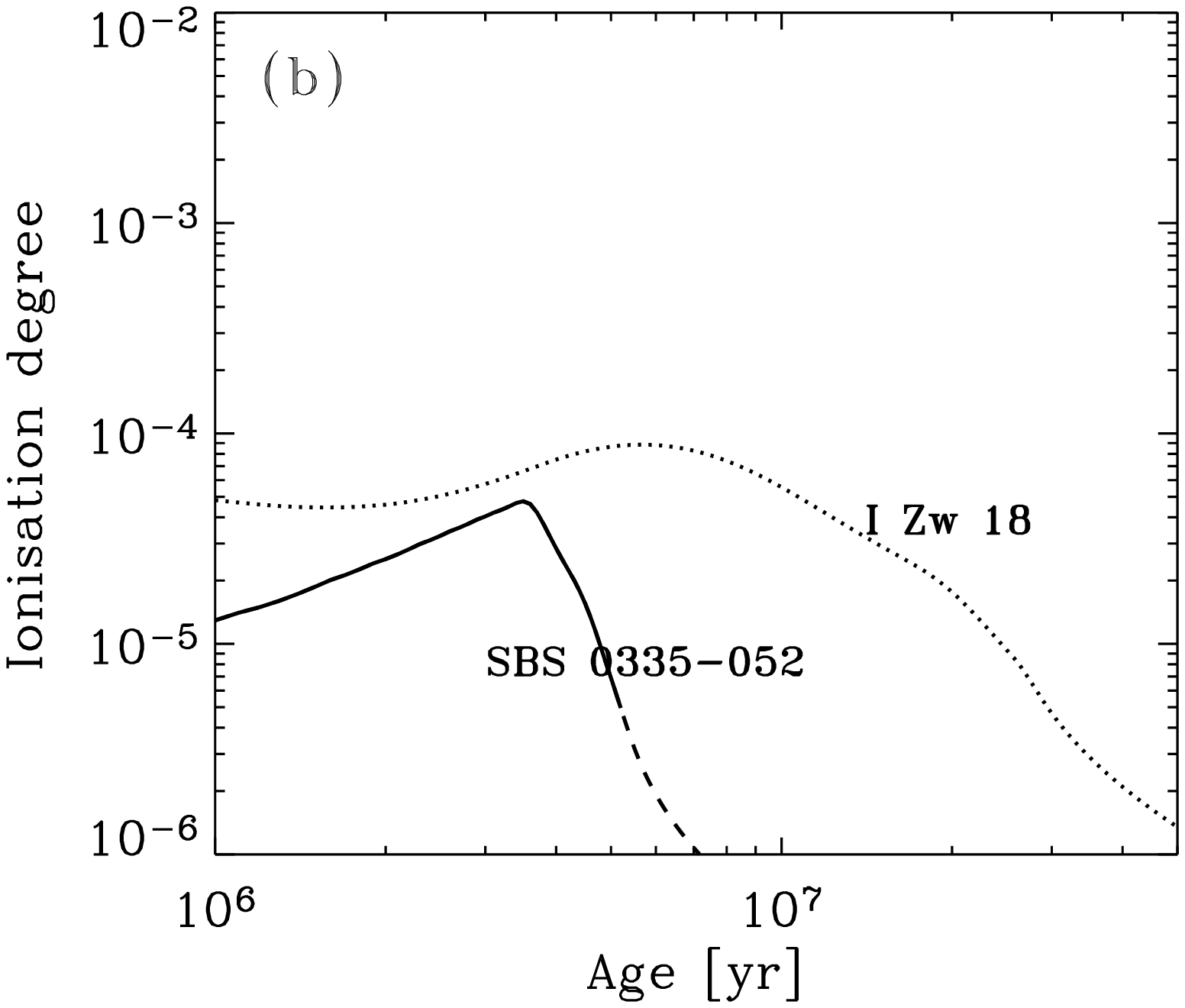}
\includegraphics[width=6cm]{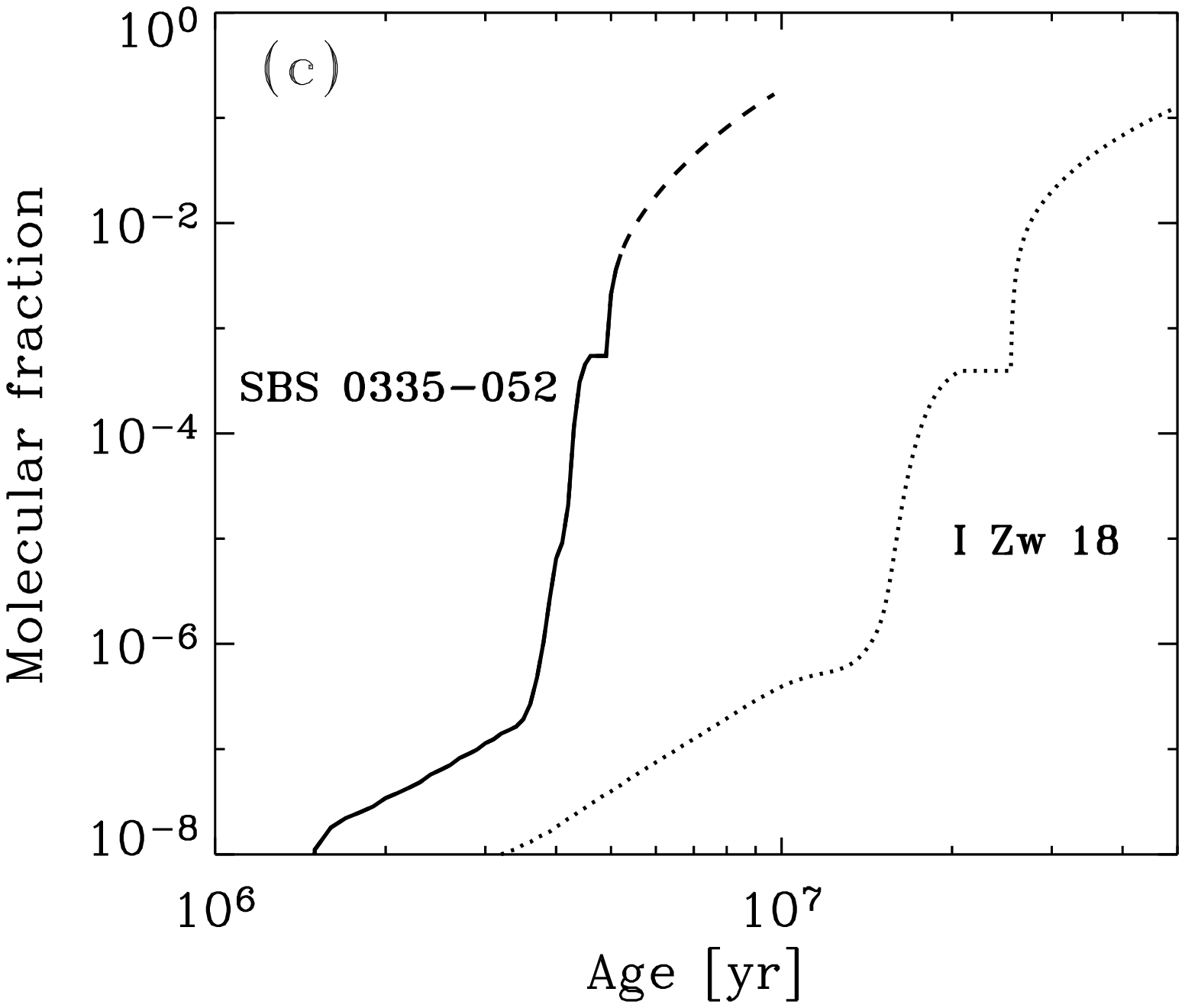}
\caption{Time evolution of {\bf a)} gas temperature,
{\bf b)} ionisation degree, and {\bf c)} molecular fraction
for the model of \sbs\ and \izw\ (solid and dotted lines,
respectively).
The dashed line continuing from the solid line shows the
evolution after $t_{\rm gas}/2$, when more than half of the
initial gas mass is converted into stars.}
\label{fig:izw}
\end{figure*}

We adopt $r_{\rm SF}=100$ pc and $n=110$ ${\rm cm}^{-3}$
(Izotov et al.\ \cite{izotovetal99}; HHTIV).
Eq.~(\ref{eq:density}) indicates
$M_{\rm gas}\simeq 1\times 10^7~M_\odot$ which is close to the
$M_{\rm gas}$ estimated by van Zee et al.\ (\cite{vanzee98}).
The free-fall time become $\tff\sim 10$ Myr
(Eq.\ \ref{eq:freefall}).
The observational star formation rate is
$\sim 0.04$--0.1 $M_\odot~{\rm yr}^{-1}$
(e.g., Hopkins et al.\ \cite{hopkins02}).
This range of star formation rate is consistent with
Eq.~(\ref{eq:sfr}) if we assume $\epsilon =0.04$--0.1;
we adopt $\epsilon =0.07$ as a rough central value. 
In Fig.~\ref{fig:izw}, we show the evolution of various
physical quantities typical of \izw\ by
dotted lines. The shielding of UV photons by dust
at $t\ga 20$ Myr causes the efficient cooling, recombination,
and molecular formation. In particular, if the age of \izw\
is $\la 20$ Myr, the poor molecular content is consistent
with observations (e.g.,
Vidal-Madjar et al.\ \cite{vidal00}).

Kamaya \& Hirashita (\cite{kamaya01}) suggest that \H2 formation
in \izw\ may be regulated by the gas phase reaction, rather than by
the formation on dust surfaces.
If we assume the age of the youngest burst in \izw\ to be
$\sim 10$ Myr, our calculation gives an ionisation  fraction
of $\sim 10^{-3}$, roughly consistent with
Kamaya \& Hirashita (\cite{kamaya01}). The molecular
fraction is $\sim 10^{-6}$, consistent also with the
upper limit by Vidal-Madjar et al.\ (\cite{vidal00}).

Recchi et al.\ (\cite{recchi02}) have proposed two
instantaneous bursts separated by a quiescent
period for \izw. The age of the recent burst
deduced by them (4--15 Myr) is in the range treated in
our paper, but an older burst which occurred $\sim 300$ Myr
is probably also necessary (see also Aloisi et al.\ \cite{aloisi99};
\"{O}stlin \cite{ostlin00}). Hunt et al.\ (\cite{hti03}) also
derive a young burst age of 3--15\,Myr, and a previous episode of
star formation no older 
than $\la 500$ Myr.  However, the stars with the age
$\la 10$ Myr contribute to the gas heating more than
the old stars, and therefore, the age in our model
should be taken as the age of the more recent burst.

\begin{figure*}
\includegraphics[width=9cm]{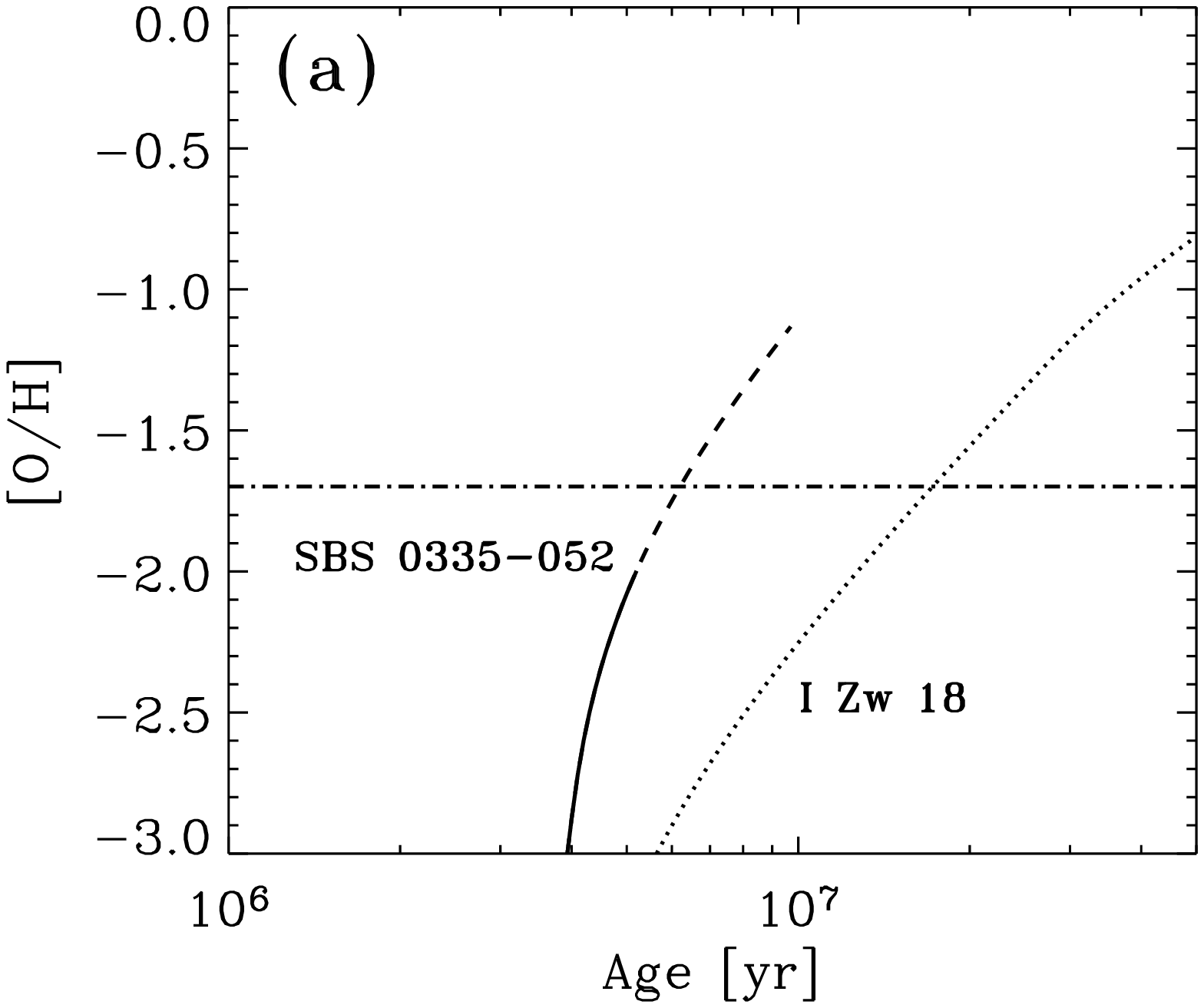}
\includegraphics[width=9cm]{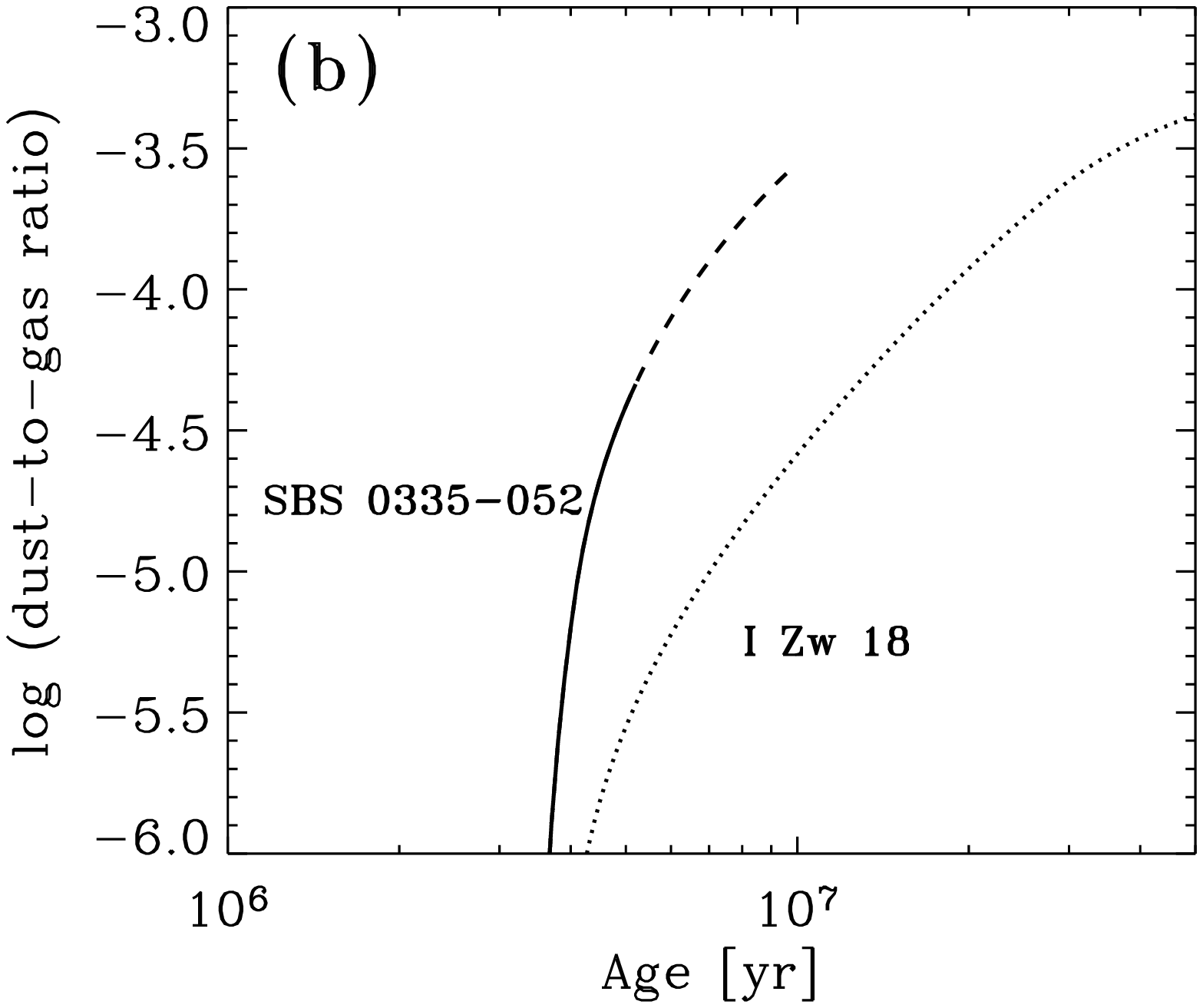}
\caption{Time evolution of {\bf a)} metallicity and
{\bf b)} dust-to-gas ratio for the model of \sbs\ and \izw\
(solid and dotted
lines, respectively). The dashed line continuing from the solid
line shows the evolution after $t_{\rm gas}$. The dash-dotted
line represents the metallicity level of 1/50 solar.}
\label{fig:izw_metal}
\end{figure*}

In Fig.~\ref{fig:izw_metal}, we present the evolution of
oxygen abundance and dust-to-gas ratio for the model of
\izw\ (dotted lines). The horizontal dash-dotted line
represents the metallicity level observed for \izw\
(1/50 solar). This metallicity is reached at the age of
$\sim 20$ Myr, which is roughly consistent with the above
ages for the current star formation activity. At this age,
the dust-to-gas ratio becomes $10^{-4}$. The observational
constraint on the dust-to-gas ratio is difficult at the
moment, since we must specify the
gas mass contained in star-forming regions, which are
generally small for BCDs, and thus difficult to resolve. 
Nevertheless, we estimate for \izw\ 
$M_{\rm gas}\simeq 1\times 10^7~M_\odot$, on the same
order of the $2.7\times 10^7~M_\odot$ found by
van Zee et al. (\cite{vanzee98}).
The dust mass of \izw\ is in the range of
2--$5\times 10^3~M_\odot$
(Cannon et al.\ \cite{cannon02}). Therefore, we obtain
a dust-to-gas ratio $\sim 10^{-4}$, which
is roughly consistent with the model calculation at the
age of $\sim 20$ Myr.

\subsubsection{\sbs}

The gas number density is roughly
$n\sim 10^3~{\rm cm}^{-3}$ (Izotov et al.\ \cite{izotovetal99};
Hunt et al.\ \cite{hunt-vla}), and the radius is
$r_{\rm SF}\sim 40$ pc. Eq.~(\ref{eq:density}) indicates that
$M_{\rm gas}\sim 10^7~M_\odot$. The free-fall time is estimated
to be $\tff\sim 3$ Myr (Eq.~\ref{eq:freefall}). In order to be
consistent with our previous paper, HHF02 ($\psi =1~M_\odot$), we
adopt a slightly large star formation efficiency,
$\epsilon =0.3$. Such a high star formation efficiency can be
correct if a region is affected by the
external pressure (Elmegreen \& Efremov \cite{elmegreen97}).

In Fig.~\ref{fig:izw} (solid line), we show the evolution of
gas temperature, ionisation degree, and molecular fraction
for \sbs\ with solid lines. The gas finally cools on a
timescale comparable to
the observationally derived young age
($\sim 5$ Myr): the temperature drops to $\la 300$ K, the
medium is kept neutral, and the molecular fraction increases.
As mentioned in Section \ref{subsec:sfr},
the calculation is stopped at $t=t_{\rm gas}/2$.
However, since the gas state rapidly changes around
$t=t_{\rm gas}/2=5$ Myr, it is interesting to show the
evolution after $t=t_{\rm gas}/2$, although 
neglecting the gas consumption becomes a bad approximation.
The dotted lines continuing after the solid lines show the
evolutions for $t>t_{\rm gas}/2$.

The evolutions of oxygen abundance and dust-to-gas ratio
is also shown in Fig.~\ref{fig:izw_metal} (solid
line). The dashed lines shows the evolution after
$t_{\rm gas}/2$ (same as Fig.~\ref{fig:izw}). We see
that the metallicity reaches 1/50 solar (roughly the
observational value 1/41 solar) around $t\sim 6$ Myr,
which is marginally consistent with the age constraint
by Vanzi et al.\ (\cite{vanzi00}) ($\la 5$ Myr).
We should note that this metallicity level is reached
after $t_{\rm gas}/2$, which implies that a significant
fraction of gas in the star-forming region has been
consumed. The run-away collapse of cooled gas could
lead to such a rapid gas consumption. The dust-to-gas
ratio is $\sim 10^{-4}$ at $t\sim 6$ Myr.

We have adopted a gas mass $M_{\rm gas}\simeq 10^7~M_\odot$
for the star-forming region, much smaller than the
observational estimate $\sim 10^9M_\odot$ by
Pustilnik et al.(\cite{pustilnik01}) for the whole galaxy.
However,
because of the lack of resolution for H {\sc i}
observation, it is difficult to put a constraint on the
gas mass in the star-forming region.
There is also a large uncertainty in dust mass.
Dale et al.\ (\cite{dale01}) estimate
$M_{\rm dust}$ to be $2400~M_\odot$ but they neglect the
stochastically
heated dust grains (in this sense, the dust mass is
underestimated), while Hunt et al.\ (\cite{hunt01})
derive $10^5~M_\odot$, but they argue that this is an upper
limit because of the assumed spatial filling factor of unity.
{}From a radiative transfer model, which however neglects
the effects of small grains,
Plante \& Sauvage (\cite{plante02}) find a dust mass of
$1.5\times10^5\,M_\odot$.
Takeuchi et al.\ (\cite{takeuchi03}) adopt the
evolution model of HHF02, and they explain the infrared
spectrum of \sbs\ with
$M_{\rm dust}\sim 5\times 10^3~M_\odot$, which
is smaller than the upper limit by Hunt et al.\
(\cite{hunt01}) but is larger than the possibly underestimated
value by Dale et al.\ (\cite{dale01}).

By using $M_{\rm dust}\sim 2\times 10^3$--$10^5$ $M_\odot$ and
$M_{\rm gas}\sim 10^7~M_\odot$, we obtain a range of the
dust-to-gas ratio in \sbs\ of $2\times 10^{-4}$--$10^{-2}$,
which is larger than that expected from our model. Because of
the uncertainties in the dust mass and the gas mass, further
observations are needed to better constrain these
numbers for \sbs. Recently, Inoue (\cite{inoue03}) also
has discussed the dust-to-gas ratio of \sbs\ with his model
but he has reproduced the observational dust-to-gas ratio
with $t\ga 10^7$ yr. Nozawa et al.\ (\cite{nozawa03})
have proposed a SN dust production larger than Todini \&
Ferrara (\cite{todini01}), and their model could solve the
discrepancy between theory and observation. 

\subsection{Possible intermediate active/passive BCDs}

The active and passive modes of star formation
are clearly two extremes on a continuum.
In this paper, we have singled out two parameters that
determine a galaxy's position on this continuum,
namely size $r_{SF}$ and density $n$. 
Dust properties distinguish the two extremes, but they
also must be viewed as part of a continuum of possibilities. 

Figure 1 of HHTIV indicates that active and passive BCDs
are clearly separated into two sequences in the surface
brightness -- density relation. However, there are 
BCDs which could be considered as intermediate between 
the active and passive extrema. For example,
Mrk\,71 (NGC\,2366), whose metallicity is about 1/10 $Z_\odot$
(Peimbert et al.\ \cite{peimbert86}), has a compact
star-forming region
but a low surface brightness (Hunt et al., in preparation).
Moreover, no SSCs are found in this galaxy by
Billett et al.\ (\cite{billett02}). Another possible
intermediate galaxy is Tol 1214-277, whose metallicity
is about 1/25 $Z_\odot$ (Izotov et al.\ \cite{izotov01}).
It has a dense ($n\simeq 210$ cm$^{-3}$;
Fricke et al.\ \cite{fricke01}), but rather diffuse
star-forming region ($\mbox{size}\sim 200$ pc;
Hunt et al., in preparation). However, since
this galaxy is distant (104 Mpc), it is difficult to constrain
the size star-forming region of Tol 1214-277 even with
the spatial resolution of {\it Hubble Space Telescope}.

\section{Implications for high redshift \label{sec:highz}}

Large numbers of primeval galaxies are expected to exist in
the high-$z$ universe ($z\ga 5$; e.g.,
Scannapieco et al.\ \cite{evan03}).
Because such galaxies must necessarily be chemically unevolved,
it is useful to extend our concept of the active-passive
dichotomy to high $z$. The ``active'' mode of star
formation is particularly relevant for high-$z$ galaxies,
because of high gas density $\ga 1000~{\rm cm}^{-3}$
(Norman \& Spaans \cite{norman97}) and warm dust content.

Galaxies become IR luminous on much shorter timescales than
the cosmic age at $z\sim 5$ (HF02), even at metallicities is as
low as \izw\ or \sbs\
(see also Morgan \& Edmunds \cite{morgan03}).
We have shown that during an ``active'' star formation
episode with no prior chemical enrichment,
dust produced by SNe II efficiently shields the UV
radiation. This means that the heating by UV photons
in the ISM is not sufficient to halt gas cooling.
Therefore, it is important to trace the high-$z$
star formation activity in the IR (or sub-mm)
(e.g., Chapman et al. \cite{chapman01}). Indeed, large
extinction has been shown by some samples of high-$z$ galaxies
(e.g., Meurer et al.\ \cite{meurer99}; 
Stanway et al. \cite{stanway03}).

The ``active'' mode is also characterised by high dust
temperature ($\ga 40$ K) because of the high UV interstellar
radiation field. Totani \& Takeuchi (\cite{totani02}) have
shown that the existence of high temperature dust at
$z\sim 3$ is favoured to explain the far-IR cosmic background
radiation.
Thus, it is important to investigate a potential increase of 
the ``active'' star formation mode at high $z$ (especially the 
increase of a population with high dust temperature), in order 
to quantify the contribution of such galaxies to the cosmic 
IR background.

Indeed, virtually all high-redshift star formation may occur
in the ``active'' regime.
It is not clear whether there are high $z$ objects corresponding to the
``passive'' mode of star formation. \H2 has been difficult to
detect from damped Ly$\alpha$ clouds (DLAs;
e.g., Petitjean et al.\ \cite{petitjean02};
Ledoux et al.\ \cite{ledoux03}). In general, stringent
upper limits of $\la 10^{-6}$ have been placed on the
molecular fractions for DLAs. In passive star-forming
regions, the molecular fraction is kept as low as
$\sim 10^{-5}$ on timescales of several tens of Myr.
Therefore, some of the DLAs could be ``passive''
star-forming objects. However, 
the low abundance of molecules can also be explained by
photodissociation by the cosmic UV background
radiation (e.g., Hirashita et al.\ \cite{hirashita03}).

\section{Conclusions}\label{sec:conclusion}

In this paper, we have theoretically modeled the ``active'' 
and ``passive'' star formation modes observed in metal-poor
BCDs. The ``active'' mode is characterised by dense and compact
($n\ga 500~{\rm cm}^{-3}$ and $r_{\rm SF}\la 50$ pc)
star-forming regions. On the contrary, the ``passive''
mode takes place in diffuse ($n\la 100~{\rm cm}^{-3}$
and $r_{\rm SF}\ga 100$ pc) star-forming regions.

In the dense regions observed in ``active'' BCDs, the
gas free-fall time is typically shorter than $\sim 5$
Myr. Such a short free-fall time can enhance
star formation activity and generate an efficient
supply of dust from SNe II.
An accumulation of dust in such a compact region
leads to a large dust optical depth, and
the region consequently becomes luminous in the 
IR. Even in the environment of active star formation,
the gas retains the physical conditions favourable
for the gas collapse: cool ($\la 1000$ K) and
highly molecular ($\fH2\ga 10^{-2}$).
The above characteristics explain the properties
of star-forming regions in the BCDs categorised as
``active'' in HHTIV, 
especially SBS 0335$-$052 (i.e., IR
luminous, containing hot dust, rich in molecules,
compact, and dense).

In ``passive'' BCDs, the dynamical time of diffuse regions
is longer than $10^7$ yr, and the star formation rate
can be as low as $\la 0.1~M_\odot$. The increase of dust
optical depth is milder and therefore ``passive''
star-forming regions become IR luminous much later in their
evolution, $> \ 10^7$ yr. Such a passive region has a low $\la 10^{-4}$
molecular content, and it would be difficult to detect \H2
in passive BCDs.

However, the physical state of gas is strongly affected
by the size and spatial distribution of grains. With a
fixed total dust mass, if the grain radius is large,
the optical depth of dust against the UV light is small.
This means for large grains ($\sim$0.01~$\mu$m), 
the gas temperature and ionisation degree tend to be high,
and the molecular fraction tends to be low. As for the
spatial distribution of grains, efficient shielding of
UV light takes place in a screen geometry, which leads
to efficient cooling and molecule formation. On the
contrary, in the mixed geometry, because of inefficient
shielding, the gas is heated to nearly $10^4$ K, and
the molecules are efficiently dissociated.

We also discussed
the evolution of dust-to-gas ratio and metallicity. 
In particular, our model is consistent with
the observations of \izw\ and \sbs, but
future observations of dust in metal-poor
(${\rm [O/H]}\la -1$) BCDs are needed for further
constraints. The consistency also implies that around
10--20 \% of metals supplied from SNe are in the dust
phase.

We have shown how differences in two physical parameters of a star-forming
region, its size and density, can lead to substantially
different evolution over time.
The distinction has been made between
``active'' and ``passive'' modes, but such a dichotomy
is perhaps misleading, since they are rather extremes
on a continuum.
Pressure must also drive evolution of a star-forming region, 
and may determine a region's initial size and density.
However, after the onset of star formation, dust
shielding of UV photons determines the fate of
star-forming regions, which finally bifurcates into
active and passive regimes.

\begin{acknowledgements}

We thank the anonymous referee for useful comments which improved
this paper considerably. We are grateful to T. X. Thuan, A. Ferrara,
M. A. R. Kobayashi, and H. Shibai for stimulating discussions on
galaxy evolution. H. H. was supported by
the Research Fellowship of the Japan Society for the Promotion of
Science for Young Scientists.  We fully utilised the NASA's
Astrophysics Data System Abstract Service (ADS).

\end{acknowledgements}

\appendix

\section{Equilibrium dust temperature}\label{app:T_dust}

We derive and discuss the equilibrium dust temperature in
Eq.~(\ref{eq:T_dust_eq}). We start from the following
equation between absorbed and emitted energy of a grain:
\begin{eqnarray}
cu_{\rm UV}\pi a^2Q_{\rm UV}=4\pi a^2\int_0^\infty
Q_{\rm IR}(a,\,\lambda)\pi B_\lambda (T_{\rm dust})\, d
\lambda\, ,
\end{eqnarray}
where $u_{\rm UV}$ is the radiative energy in UV defined
in Eq.~(\ref{eq:isrf}), $a$ is the grain radius,
$B_\lambda (T_{\rm dust})$ is the Planck function estimated at
the wavelength of $\lambda$ and the dust temperature of
$T_{\rm dust}$, and 
$Q_{\rm UV}$ and $Q_{\rm IR}(a,\,\lambda)$ are the absorption
cross sections of a grain normalized by the geometrical cross
section in UV and IR, respectively. We assume that $Q_{\rm UV}$
is independent of $\lambda$ (Draine \& Lee \cite{draine84}),
and adopt the following form for $Q_{\rm IR}(a,\,\lambda)$:
\begin{eqnarray}
Q_{\rm IR}(a,\,\lambda )=\frac{2\pi Aa}{\lambda^\beta}\, ,
\end{eqnarray}
In the main text, we assume that $\beta =2$
(Drapatz \& Michel \cite{drapatz77};
Shibai et al.\ \cite{shibai99};
Takeuchi et al.\ \cite{takeuchi03}), but we generalise the
wavelength dependence.
Since the Planck function is written as
\begin{eqnarray}
B_\lambda (T)=\frac{2hc^2/\lambda^5}{\exp (hc/\lambda kT)-1}\,
,
\end{eqnarray}
where $k$ is the Boltzmann constant, we obtain the following
analytic solution for $T_{\rm dust}$:
\begin{eqnarray}
T_{\rm dust}=\frac{hc}{k}\left(
\frac{u_{\rm UV}Q_{\rm UV}}{16\pi^2AahcI(3+\beta )}
\right)^{1/(4+\beta)}\, ,
\end{eqnarray}
where the function $I(\alpha )$ is defined as
\begin{eqnarray}
I(\alpha )\equiv \int_0^\infty\frac{x^\alpha}{e^x-1}\, dx
~~~(\alpha >0).
\end{eqnarray}
Since $I(5)=8\pi^6/63$, we obtain Eq.~(\ref{eq:T_dust_eq})
for $\beta =2$.



\begin{thebibliography}{}
\bibitem[2000]{abel00} Abel, T., Bryan, G. L., \& Norman, M. L.
    2000, ApJ, 540, 39
\bibitem[1999]{aloisi99} Aloisi, A., Tosi, M., \& Greggio, L.
    1999, AJ, 118, 302
\bibitem[2001]{alton01} Alton, P. B., Lequeux, J., Bianchi, S.,
    et al.\ 2001, A\&A, 366, 451
\bibitem[2002]{barkana02} Barkana, R. 2002, NewA, 7, 85
\bibitem[2001]{bekky01} Bekki, K., \& Couch, W. J. 2001, ApJ,
    557, L19
\bibitem[2002]{bendo02} Bendo, G. J., Joseph, R. D.,
    Wells, M., et al.\ 2002, AJ, 124, 1380
\bibitem[1999]{bianchi99} Bianchi, S., Davies, J. I., \&
    Alton, P. B. 1999, A\&A, 344, L1
\bibitem[2002]{billett02} Billett, O. H., Hunter, D. A., \&
    Elmegreen, B. G. 2002, AJ, 123, 1454
\bibitem[1956]{bonnor56} Bonnor, W. B. 1956, MNRAS, 116, 351
\bibitem[1988]{boulanger88} Boulanger, F., Beichman, C.,
    D\'{e}sert, F. X., Helou, G., P\'{e}rault, M., \&
    Ryter, C. 1988, ApJ, 332, 328
\bibitem[2001]{bromm01} Bromm, V., Coppi, P. S., \& Larson, R. B.
    2001, ApJ, 564, 23
\bibitem[2002]{cannon02} Cannon, J. M., Skillman, E. D.,
    Garnett, D. R., \& Dufour, R. J. 2002, ApJ, 565, 931
\bibitem[2002]{cazaux02} Cazaux, S., \& Tielens, A. G. G. M. 2002,
    ApJ, 575, L29
\bibitem[2001]{chapman01} Chapman, S. C., Richards, E. A., 
Lewis, G. F., Wilson, G., \& Barger, A. J. 2001, ApJ, 548, 147 
\bibitem[2000]{ciardi00} Ciardi, B., Ferrara, A., Governato, F.,
    \& Jenkins, A. 2000, MNRAS, 314, 611
\bibitem[2000]{contursi00} Contursi, A., Lequeux, J., Cesarsky, D.,
et al. 2000, A\&A, 362, 310
\bibitem[2000]{cox00} Cox, A. N. 2000, Allen's Astrophysical
    Quantities, 4th ed.\ (Springer, New York)
\bibitem[2001]{dale01} Dale, D. A., Helou, G., Neugebauer, G.,
    Soifer, B. T., Frayer, D. T., \& Condon, J. J. 2001, AJ, 122,
    1736
\bibitem[1984]{draine84} Draine, B. T., \&  Lee, H. M.
    1984, ApJ, 285, 89
\bibitem[1977]{drapatz77} Drapatz, S., \& Michel, K. W. 1977, A\&A,
    56, 353
\bibitem[1980]{dwek80} Dwek, E., \& Scalo, J. M. 1980, ApJ, 239, 193
\bibitem[1998]{dwek98} Dwek, E. 1998, ApJ, 501, 643
\bibitem[1955]{ebert55} Ebert, R. 1955, Z. Astrophys., 37, 222
\bibitem[2000]{elmegreen00} Elmegreen, B. G. 2000, ApJ, 530, 277
\bibitem[1997]{elmegreen97} Elmegreen, B. G., \& Efremov, Y. N.
    1997, ApJ, 480, 235
\bibitem[1978]{elmegreen78} Elmegreen, B. G., \& Elmegreen, D. M.
    1978, ApJ, 220, 1051
\bibitem[2000]{eh00} Elmegreen, B. G., \& Hunter, D. A. 2000, ApJ, 540, 814
\bibitem[1986]{peimbert86} Peimbert, M., Pe\~{n}a, M., \&
    Torres-Peimbert, S. 1986, A\&A, 158, 266
\bibitem[2000]{ferrara00} Ferrara, A., Pettini, M. \&
    Shchekinov, Y. 2000, MNRAS, 319, 539
\bibitem[2001]{fricke01} Fricke, K. J., Izotov, Y. I.,
    Papaderos, P., et al.\ 2001, AJ, 121, 169
\bibitem[1998]{galli98} Galli, D., \& Palla, F. 1998, A\&A, 335,
    403
\bibitem[2003]{galliano03} Galliano, F., Madden, S. C., Jones, A. P.,
    et al.\ 2003, A\&A, 407, 159
\bibitem[2003]{habart03} Habart, E., Boulanger, F., Verstraete, L.,
    et al.\ 2003, A\&A, 397, 623
\bibitem[1983]{hildebrand83} Hildebrand, R. H. 1983, QJRAS, 24, 267
\bibitem[2002]{hirashita02} Hirashita, H., \& Ferrara, A. 2002, MNRAS,
    337, 921 (HF02)
\bibitem[2003]{hirashita03} Hirashita, H., Ferrara, A., Wada, K.,
    \& Richter, P. 2003, MNRAS, 341, L18
\bibitem[2002]{hhf02} Hirashita, H., Hunt, L. K., \& Ferrara, A.
    2002, MNRAS, 330, L19 (HHF02)
\bibitem[2002]{hopkins02} Hopkins, A. M., Schulte-Ladbeck, R. E.,
    Drozdovsky, I. O. 2002, AJ, 124, 862
\bibitem[2002]{hunt02} Hunt, L. K., Giovanardi, C., \& Helou, G.
    2002, A\&A, 394, 873
\bibitem[2004a]{hunt-vla} Hunt, L.K., Dyer, K.D.,
Thuan, T.X., Ulvestad, J.S. 2004a, ApJ, submitted
\bibitem[2004b]{hunt-active} Hunt, L. K., Hirashita, H., \& Thuan, T. X.
    2004b, in preparation
\bibitem[2003a]{hunt-cozumel} Hunt, L.K., Hirashita, H.,
Thuan, T.X., Izotov, Y.I., \& Vanzi, L. 2003a, in 
Proceedings of ``Galaxy Evolution:  Theory and Observations'',   
Eds.  V. Avila-Reese, C. Firmani,  C. Frenk, \& C. Allen, RevMexAA SC
(HHTIV) ({\it astro-ph/0310865})
\bibitem[2003b]{hti03} Hunt, L. K., Thuan, T. X., \& Izotov, Y. I.
    2003b, ApJ, 588, 281
\bibitem[2001]{hunt01} Hunt, L. K., Vanzi, L., \& Thuan, T. X. 2001,
    A\&A, 377, 66
\bibitem[1994]{hunter94} Hunter, D. A., O'Connell, R. W., \&
    Gallagher, J. S. III, 1994, AJ, 108, 84
\bibitem[2003]{inoue03} Inoue, A. K. 2003, PASJ, 55, 901
\bibitem[2000]{inoue00} Inoue, A. K., Hirashita, H., \& Kamaya, H.
    2000, AJ, 120, 2415
\bibitem[2001]{inoue01} Inoue, A. K., Hirashita, H., \& Kamaya, H.
    2000, ApJ, 555, 613
\bibitem[1999]{izotovetal99} Izotov, Y. I., Chaffee, F. H.,
    Foltz, C. B., et al.\ 1999, ApJ, 527, 757
\bibitem[2001]{izotov01} Izotov, Y. I., Chaffee, F. H., \&
    Green, R. F. 2001, ApJ, 562, 727
\bibitem[1999]{izotov99} Izotov, Y. I., \& Thuan, T. X. 1999,
    ApJ, 511, 639
\bibitem[2002]{james02} James, A., Dunne, L., Eales, S., \&
    Edmunds, M. G. 2002, MNRAS, 335, 753
\bibitem[1996]{jones96} Jones, A. P.,
    Tielens, A. G. G. M., \& Hollenbach, D. J. 1996, ApJ, 469, 740
\bibitem[2001]{kamaya01} Kamaya, H., \& Hirashita, H. 2001, PASJ,
    53, 483
\bibitem[2002]{kamaya02} Kamaya, H., \& Silk, J. 2002, MNRAS, 332,
    251
\bibitem[2000]{kitayama00} Kitayama, T., \& Ikeuchi, S. 2000, ApJ,
    529, 615
\bibitem[2001]{kitayama01} Kitayama, T., Susa, H., Umemura, M.,
    \& Ikeuchi, S. 2001, MNRAS, 326, 1353
\bibitem[1989]{kozasa89} Kozasa, T., Hasegawa, H., \& Nomoto, K.
    1989, ApJ, 344, 325
\bibitem[2000]{kunth00} Kunth, D., \& \"{O}stlin, G. 2000, A\&AR,
    10, 1
\bibitem[2003]{ledoux03} Ledoux, C., Petitjean, P., \&
    Srianand, R. 2003, MNRAS, 346, 209
\bibitem[1992]{lin92} Lin, D. N. C., \& Murray, S. D. 1992, ApJ, 394, 523 
\bibitem[1998]{lisenfeld98} Lisenfeld, U., \&
    Ferrara, A. 1998, ApJ, 496, 145 
\bibitem[1969]{matsuda69} Matsuda, T., Sato, H., \& Takeda, H. 1968,
    Prog.\ Theor.\ Phys., 42, 219
\bibitem[1999]{meurer99} Meurer, G. R., Heckman, T. M., \&
    Calzetti, D. 1999, ApJ, 521, 64
\bibitem[1989]{mckee89} McKee, C. F. 1989, in IAU Symp.\ 135,
    Interstellar Dust, ed. L. J.
    Allamandola \& A. G. G. M. Tielens (Dordrecht: Kluwer), 431
\bibitem[2003]{morgan03} Morgan, H. L., \& Edmunds, M. G. 2003,
    MNRAS, 343, 427
\bibitem[1984]{natta84} Natta, A., \& Panagia, N. 1984, ApJ, 287, 228
\bibitem[1998]{nishi98} Nishi, R., Susa, H., Uehara, H., et al.\
    1998, Prog.\ Theor.\ Phys., 100, 881
\bibitem[1997]{norman97} Norman, C. A., \& Spaans, M. 1997,
    ApJ, 480, 145
\bibitem[2003]{nozawa03} Nozawa, T., Kozasa, T., Umeda, H.,
    et al.\ 2003, ApJ, 598, 785
\bibitem[2000]{omukai00} Omukai, K. 2000, ApJ, 534, 809
\bibitem[2000]{ostlin00} \"{O}stlin, G. 2000, ApJ, 535, L99
\bibitem[1968]{peebles68} Peebles, P. J. E., \& Dicke, R. H. 1968,
    ApJ, 154, 891
\bibitem[2002]{petitjean02} Petitjean, P., Srianand, R., \&
    Ledoux, C. 2002, MNRAS, 332, 383
\bibitem[2002]{plante02} Plante, S., \& Sauvage, M. 2002, AJ, 124,
    1995
\bibitem[1999]{popescu99} Popescu, C. C., Hopp, U., \& Rosa, M. R.
    1999, A\&A, 350, 414
\bibitem[1989]{puget89} Puget, J. L., \& L{\' e}ger, A. 1989, ARA\&A,
27, 161
\bibitem[2001]{pustilnik01} Pustilnik, S.~A., 
Brinks, E., Thuan, T.~X., Lipovetsky, V.~A., \& Izotov, Y.~I.\ 2001, AJ, 
121, 1413 
\bibitem[2002]{recchi02} Recchi, S., Matteucci, F.,
    D'Erdole, A., \& Tosi, M. 2002, A\&A, 384, 799
\bibitem[2002]{ripamonti02} Ripamonti, E., Haardt, F., Ferrara, A.,
    \& Colpi, M. 2002, MNRAS, 334, 401
\bibitem[2003]{salvaterra03} Salvaterra, R., \& Ferrara, A. 2003,
    MNRAS, 339, 973
\bibitem[2003]{evan03} Scannapieco, E., Schneider, R., \&
    Ferrara, A. 2003, ApJ, 589, 35
\bibitem[2002]{schaerer02} Schaerer, D. 2002, A\&A, 382, 28
\bibitem[2004]{raffa04} Schneider, R., Ferrara, A., \&
    Salvaterra, R. 2004, MNRAS, submitted
\bibitem[1972]{searle72} Searle, L., \& Sargent, W. L. W. 1972,
    ApJ, 173, 25
\bibitem[1999]{shibai99} Shibai, H., Okumura, K., \& Onaka, T.
    1999, Star Formation 1999, ed.\ T. Nakamoto (Nobeyama:
    Nobeyama Radio Observatory), 67
\bibitem[1998]{silva98} Silva, L., Granato, G. L., Bressan, A.,
    \& Danese, L. 1998, ApJ, 509, 103
\bibitem[1993]{skillman93} Skillman, E. D., \& Kennicutt,
    R. C., Jr.\ 1993, ApJ, 411, 655
\bibitem[1997]{smail97} Smail, I., Ivison, R. J., \&
    Blain, A. W. 1997, ApJ, 490, L5
\bibitem[2003]{stanway03} Stanway, E. R., Bunker, A. J., \& McMahon, 
R. G. 2003, MNRAS, 342, 439
\bibitem[2003]{takeuchi03} Takeuchi, T. T., Hirashita, H.,
    Ishii, T. T., Hunt, L. K., \& Ferrara, A. 2003, MNRAS,
    343, 839
\bibitem[2004]{takeuchi04} Takeuchi, T. T., Yoshikawa, K., \&
    Tomita, A. 2004, ApJ, submitted
\bibitem[1997]{tegmark97} Tegmark, M., Silk, J., Rees, M. J.,
    et al.\ 1997, ApJ, 474, 1
\bibitem[1999]{thuan99} Thuan, T.~X., Sauvage, M., \& Madden, S.\
    1999, ApJ, 516, 783 
\bibitem[1998]{tielens98} Tielens, A. G. G. M. 1998, ApJ, 499, 267
\bibitem[1980]{tinsley80} Tinsley, B. M. 1980,
    Fundam.\ Cosmic Phys., 5, 287
\newpage
\bibitem[2001]{todini01} Todini, P., \& Ferrara, A. 2001, MNRAS,
    325, 726
\bibitem[2002]{tomita02} Tomita, A., Yoshikawa, K.,
    Takeuchi, T. T., Hirashita, H. 2002, in Proc.\ of the IAU
    8th Asian-Pacific Regional Meeting, Vol.\ II, ed.\
    S. Ikeuchi, J. Hearnshaw, T. Hanawa, (Tokyo: Astronomical
    Society of Japan), 301
\bibitem[2002]{totani02} Totani, T., \& Takeuchi, T. T. 2002,
    ApJ, 570, 470
\bibitem[1997]{tsujimoto97} Tsujimoto, T., Yoshii, Y., Nomoto, K.,
    et al.\ 1997, ApJ, 483, 228
\bibitem[1998]{vanzee98} van Zee, L., Westpfahl, D.,
    Haynes, M. P., \& Salzer, J. J. 1998, AJ, 115, 1000
\bibitem[2000]{vanzi00} Vanzi, L., Hunt, L. K., Thuan, T. X., \&
    Izotov, Y. I. 2000, A\&A, 363, 493
\bibitem[2000]{vidal00} Vidal-Madjar, A., Kunth, D.,
    Lecavelier des Etangs, A., et al.\ 2000, ApJ, 538, L77
\bibitem[1992]{voit92} Voit, G.M. 1992, MNRAS, 258, 841
\bibitem[2002]{walter02} Walter, F., Weiss, A.,
    Martin, C., \& Scoville, N. 2002, AJ, 123, 225
\bibitem[2000]{wilson00} Wilson, C. D., Scoville, N.,
    Madden, S. C., \& Charmandaris, V. 2000, ApJ, 542, 120
\bibitem[1995]{woosley95} Woosley, S. E., \& Weaver, T. A. 1995,
    ApJS, 101, 181
\end{thebibliography}
\end{document}